\documentclass[12pt, draftclsnofoot, onecolumn]{IEEEtran}
\usepackage[T1]{fontenc}
\usepackage{amssymb}
\usepackage[pdftex]{graphicx}
\usepackage{caption}[2020/09/12]
\usepackage{amsmath}
\usepackage{amsthm}
\usepackage{mathtools}
\usepackage{array}
\usepackage[noadjust]{cite}
\usepackage{algorithm}
\usepackage{algpseudocode}
\usepackage{multirow}
\usepackage{xcolor}
\usepackage[utf8]{inputenc}
\usepackage{textcomp}
\usepackage[hidelinks]{hyperref}
\usepackage[acronym]{glossaries}
\usepackage[binary-units=true]{siunitx}

\algnewcommand{\Initialize}[1]{%
  \State \textbf{Initialize}
  \Statex \hspace*{\algorithmicindent}\parbox[t]{.8\linewidth}{\raggedright #1}
}

\makeatletter
\let\MYcaption\@makecaption
\makeatother

\algnewcommand{\LineComment}[1]{\(\triangleright\) #1}

\DeclareSIUnit\frame{frame}
\DeclareSIUnit{\kmph}{kmph}
\DeclareSIUnit\PRB{PRB}
\DeclareMathAlphabet{\mathbcal}{OMS}{cmsy}{b}{n}

\newacronym{3GPP}{3GPP}{3rd Generation Partnership Project}
\newacronym{5GNR}{5G NR}{5G New Radio}

\newacronym{AMP}{AMP}{approximate message passing}
\newacronym{AWGN}{AWGN}{additive white Gaussian noise}

\newacronym{BER}{BER}{bit error rate}
\newacronym{BICM}{BICM}{bit-interleaved coded modulation}
\newacronym{BLER}{BLER}{block error rate}
\newacronym{BMDR}{BMDR}{bit-metric decoding rate}
\newacronym{BS}{BS}{base station}

\newacronym{CDF}{CDF}{cumulative distribution function}
\newacronym{CER}{CER}{codeword error rate}
\newacronym{CNN}{CNN}{convolutional neural network}
\newacronym{Conv2D}{Conv2D}{convolutional}
\newacronym{CSI}{CSI}{channel state information}

\newacronym{DMRS}{DMRS}{demodulation reference signal}

\newacronym{eMBB}{eMBB}{enhanced mobile broadband}
\newacronym{EP}{EP}{expectation propagation}

\newacronym{FC}{FC}{fully-connected}

\newacronym{GMI}{GMI}{generalized mutual information}

\newacronym{HARQ}{HARQ}{hybrid automatic repeat request}

\newacronym{iid}{i.i.d.\@}{independent and identically distributed}
\newacronym{ISI}{ISI}{inter-symbol interference}

\newacronym{KL}{KL}{Kullback–Leibler}

\newacronym{LA}{LA}{link-adaptation}
\newacronym{LDPC}{LDPC}{low-density parity-check}
\newacronym{LMMSE}{LMMSE}{linear minimum mean square error}
\newacronym{LLR}{LLR}{log-likelihood ratio}

\newacronym{MAP}{MAP}{maximum a posteriori}
\newacronym{MCS}{MCS}{modulation and coding scheme}
\newacronym{MI}{MI}{mutual-information}
\newacronym{MIMO}{MIMO}{multiple-input multiple-output}
\newacronym{ML}{ML}{maximum-likelihood}
\newacronym{MU-MIMO}{MU-MIMO}{multi-user MIMO}
\newacronym{MSE}{MSE}{mean squared error}

\newacronym{NN}{NN}{neural network}

\newacronym{OFDM}{OFDM}{orthogonal frequency division multiplexing}

\newacronym{PHY}{PHY}{physical layer}
\newacronym{PRB}{PRB}{physical resource block}

\newacronym{QAM}{QAM}{quadrature amplitude modulation}
\newacronym{QPSK}{QPSK}{quadrature phase-shift keying}

\newacronym{RE}{RE}{resource element}

\newacronym{SD}{SD}{standard deviation}
\newacronym{SGD}{SGD}{stochastic gradient descent}
\newacronym{SIC}{SIC}{soft interference cancellation}
\newacronym{SINR}{SINR}{signal-to-interference-noise ratio}
\newacronym{SISO}{SISO}{single-input single-ouput}
\newacronym{SLS}{SLS}{system-level simulations}
\newacronym{SNR}{SNR}{signal-to-noise ratio}
\newacronym{SRS}{SRS}{sounding reference signal}
\newacronym{SU-MIMO}{SU-MIMO}{single-user MIMO}
\newacronym{SU-SISO}{SU-SISO}{single-user SISO}

\newacronym{wrt}{w.r.t.\@}{with respect to}

\newacronym{ZF}{ZF}{zero-forcing}
\newacronym{RL}{RL}{reinforcement-learning}

\newacronym{UE}{UE}{user-equipment}
\newacronym{uRLLC}{uRLLC}{ultra-reliable low-latency communications}

\DeclareMathAlphabet{\mathbcal}{OMS}{cmsy}{b}{n}
\renewcommand{\vec}[1]{\mathbf{#1}}
\newcommand{\vecr}[1]{\mathrm{#1}}

\algblock{Input}{EndInput}
\algnotext{EndInput}
\algblock{Output}{EndOutput}
\algnotext{EndOutput}
\newcommand{\Desc}[2]{\State \makebox[3em][l]{#1}#2}

\newcommand{\be}{\begin{equation}}
\newcommand{\ee}{\end{equation}}			
\newcommand{\ben}{\begin{equation*}}
\newcommand{\een}{\end{equation*}}

\newtheorem{definition}{Definition}[section]
\newtheorem{thm}{Theorem}

\newtheorem{note}{Note}
\newtheorem{remark}{Remark}

\newcommand{\nv}{\vec{n}}

\newcommand{\sv}{\vec{s}}

\newcommand{\xv}{\vec{x}}
\newcommand{\yv}{\vec{y}}

\newcommand{\Dm}{\vec{D}}

\newcommand{\Hm}{\vec{H}}
\newcommand{\Id}{\vec{I}}

\newcommand{\Km}{\vec{K}}

\newcommand{\Qm}{\vec{Q}}
\newcommand{\Rm}{\vec{R}}

\newcommand{\Um}{\vec{U}}

\newcommand{\Wm}{\vec{W}}
\newcommand{\Xm}{\vec{X}}

\newcommand{\Hrm}{\vecr{H}}
\newcommand{\Drm}{\vecr{D}}
\newcommand{\Rrm}{\vecr{R}}
\newcommand{\yrv}{\vecr{y}}
\newcommand{\hrv}{\vecr{h}}
\newcommand{\srv}{\vecr{s}}
\newcommand{\nrv}{\vecr{n}}

\newcommand{\Bc}{{\cal B}}
\newcommand{\Cc}{{\mathcal{C}}}
\newcommand{\Dc}{{\cal D}}
\newcommand{\Ec}{{\cal E}}

\newcommand{\Gc}{{\cal G}}
\newcommand{\Hc}{{\cal H}}

\newcommand{\Lc}{{\cal L}}

\newcommand{\Nc}{{\mathcal{N}}}

\newcommand{\Qc}{{\cal Q}}
\newcommand{\Rc}{{\cal R}}
\newcommand{\Sc}{{\cal S}}

\newcommand{\Uc}{{\cal U}}
\newcommand{\Wc}{{\cal W}}

\newcommand{\Xc}{{\cal X}}
\newcommand{\Yc}{{\cal Y}}

\newcommand{\Hbc}{{\mathbcal H}}
\newcommand{\Bbc}{{\mathbcal B}}
\newcommand{\Ybc}{{\mathbcal Y}}
\newcommand{\Zbc}{{\mathbcal Z}}

\newcommand{\Cbb}{\mathbb{C}}
\newcommand{\Dbb}{\mathbb{D}}

\newcommand{\Rbb}{\mathbb{R}}

\newcommand{\Zbb}{\mathbb{Z}}

\newcommand{\LB}{\left(}
\newcommand{\RB}{\right)}
\newcommand{\LP}{\left\{}
\newcommand{\RP}{\right\}}
\newcommand{\LSB}{\left[}
\newcommand{\RSB}{\right]}

\renewcommand{\ln}[1]{\mathop{\mathrm{ln}}\LB #1\RB}
\renewcommand{\log}[1]{\mathop{\mathrm{log}_2}\LB #1\RB}

\renewcommand{\exp}[1]{\mathop{\mathrm{exp}}\LB #1\RB}
\newcommand{\exps}[1]{\mathop{\mathrm{exp}}\LSB #1\RSB}

\newcommand{\EE}{{\mathbb{E}}}
\newcommand{\Expect}[2]{\EE_{#1}\LSB #2\RSB}
\newcommand{\PP}{{\mathbb{P}}}
\newcommand{\Prob}[1]{\PP \LB #1 \RB}

\newcommand{\ReLU}{\mathop{\mathrm{ReLU}}}

 \newcommand{\argmax}[1]{\underset{#1}{\operatorname{arg}\,\operatorname{max}}\;}

\begin{document}

\title{Bit-Metric Decoding Rate in Multi-User MIMO Systems: Theory}

\author{K. Pavan Srinath and Jakob Hoydis, \textit{Senior Member, IEEE}%

\thanks{K. P. Srinath is with Nokia Bell Labs, 91620 Nozay, France (email: pavan.koteshwar\_srinath@nokia-bell-labs.com), and J. Hoydis is with NVIDIA, 06906 Sophia Antipolis, France (email: jhoydis@nvidia.com). A significant part of this work was done when J. Hoydis was at Nokia Bell Labs, 91620 Nozay, France.}
}

\maketitle

\begin{abstract}
\Gls{LA} is one of the most important aspects of wireless communications where the \gls{MCS} used by the transmitter is adapted to the channel conditions in order to meet a certain target error-rate. In a \gls{SU-SISO} system with out-of-cell interference, \gls{LA} is performed by computing the post-equalization \gls{SINR} at the receiver. The same technique can be employed in \gls{MU-MIMO} receivers that use linear detectors. Another important use of post-equalization \gls{SINR} is for \gls{PHY} abstraction, where several \gls{PHY} blocks like the channel encoder, the detector, and the channel decoder are replaced by an abstraction model in order to speed up system-level simulations. However, for \gls{MU-MIMO} systems with non-linear receivers, there is no known equivalent of post-equalization \gls{SINR} which makes both \gls{LA} and \gls{PHY} abstraction extremely challenging. This important issue is addressed in this two-part paper.  In this part, a metric called the \gls{BMDR} of a detector, which is the proposed equivalent of post-equalization \gls{SINR}, is presented. Since BMDR does not have a closed form expression that would enable its instantaneous calculation, a machine-learning approach to predict it is presented along with extensive simulation results. 


\end{abstract}

\glsresetall
\section{Introduction}
\label{sec:intro}

Next-generation wireless technologies such as \gls{5GNR} are designed to deliver high levels of performance and efficiency that enable a wide range of 5G services like \gls{eMBB} and \gls{uRLLC}. Advanced \gls{MIMO} transmission techniques, wideband \gls{OFDM}, and strong channel coding schemes are some important features of these technologies. One of the most difficult challenges in a \gls{MIMO} system is the joint detection of the signal components. The task of a MIMO detector is to generate soft-information, usually in the form of \gls{LLR}, for each transmitted bit of each user. This soft-information is used by the channel decoder to recover the transmitted message bits. There exist many \gls{MIMO} detection techniques in the literature, ranging from simple linear detectors like the \gls{LMMSE} detector~\cite[Ch. 8]{Tse2005}, \cite{Cioffi1995}, to the more computationally-expensive \gls{ML}-based sphere-decoder~\cite{Viterbo1999, Studer2010}. There are several detectors whose computational complexity and reliability lie in between that of the \gls{LMMSE} detector and the sphere-decoder, a couple of them being the fixed-complexity sphere-decoder~\cite{Barbero2008} and the $K$-best detector~\cite{Guo2006}. Other non-linear detectors that are not based on sphere-decoding include iterative \gls{SIC}~\cite{SIC}, \gls{AMP}-based MIMO detectors~\cite{Jeon2015}, \gls{EP}-based~\cite{Tan2020} \gls{MIMO} detectors, and machine-learning-based receivers (see~\cite{honkala2021deeprx}, and references therein). While the \gls{LMMSE} detector is quite popular due to its low complexity, non-linear detectors will likely play an increasingly important role in next-generation MIMO systems.

\subsection{Motivation}
The following two important use cases in advanced wireless communication systems serve as motivation for this work.

\noindent 1) {\it \Gls{LA}}: One of the most important aspects of high-rate and reliable wireless communication systems is \gls{LA}. This feature enables a transmitter to adapt its \gls{MCS} to the current channel conditions in order to meet a certain target \gls{CER} or \gls{BLER}, where a block can consist of several codewords~\cite[Section 5]{3GPP_coding_2020}. This allows the transmitter and the receiver to reliably communicate at near-optimal data-rates that the channel supports. 

In \gls{MU-MIMO} transmission on the uplink, multiple users can be co-scheduled for transmission on the same set of resources, thereby potentially interfering with one another. The goal of \gls{LA} is to predict the highest supportable \gls{MCS} level for each user based only on the composite \gls{MU-MIMO} channel estimates of all the co-scheduled users before the transmission of their respective input signals. In \gls{5GNR}, these composite \gls{MU-MIMO} channel estimates can be obtained from the individual channel estimates of each user either from the periodic \gls{SRS} sent by each user, or from the previous estimates of the individual user's channel by the base station. There is no guarantee that the same set of users will be co-scheduled in multiple time slots, so it would be difficult to estimate the MCS level of a user based on its previously used MCS level or its previously seen \gls{SINR}. To illustrate with an example, suppose that users $A$, $B$, and $C$ were co-scheduled in the previous transmission slot, and that $A,B,D$ are to be co-scheduled in the next slot. Then, the interference seen by $A$ and $B$ would be different in the two cases unless the users' channels are mutually orthogonal, which is rarely the case. Also, if $D$ is transmitting for the first time, its potential \gls{SINR} when paired with $A$ and $B$ would be unknown. When users are co-scheduled and interfere with one another, post-equalization \gls{SINR} cannot be obtained from pilot signals because pilots of different users are designed to be mutually orthogonal while the data signals are not. For a linear detector, one can obtain a post-equalization \gls{SINR} estimate for each user using textbook formula~\cite[Chapter 3]{studer2009iterative} based only on the composite \gls{MU-MIMO} channel estimate and the thermal noise variance (which can always be estimated). This estimated post-equalization \gls{SINR} is a sufficient metric to identify the most suitable \gls{MCS}~\cite{la_2012,la_2020}. The relevance of post-equalization \gls{SINR} in a \gls{MU-MIMO} setup is in the fact that it can be rigorously related to achievable rates via outage capacity lower bounds. In the example considered, the base station can form a composite \gls{MU-MIMO} channel estimate $\hat{\Hm} = \LSB \hat{\Hm}_A, \hat{\Hm}_B, \hat{\Hm}_D\RSB$ for users $A,B$, and $D$ based on the previous channel estimates $\hat{\Hm}_A$ and $\hat{\Hm}_B$ of $A$ and $B$, respectively, and an estimate $\hat{\Hm}_D$ of user $D$'s channel from its transmitted \gls{SRS}. However, for non-linear detectors including those based on approximate message passing~\cite{Jeon2015} and expectation propagation~\cite{Tan2020}, there is no known method in the literature to obtain the equivalent of post-equalization \gls{SINR} based only on the estimated \gls{MU-MIMO} channel. For these iterative algorithms, one needs the composite received data signal from all the co-scheduled users in order to generate soft-symbols and an estimated \gls{SINR}, which is not possible using pilot transmissions. For sphere-decoding and its variants, it is not yet known how to obtain such a post-equalization \gls{SINR} estimate even after reception of data signals, making \gls{LA} very challenging. 

\noindent 2) {\it \Gls{PHY} abstraction}: \Gls{BS} and \gls{UE} manufacturers use \gls{SLS} to evaluate the performance of their algorithms. A typical system-level simulator is composed of (but not limited to) the following functionalities: intercell-interference modelling, resource scheduling and resource allocation, power allocation and power control, link-adaptation block with channel quality feedback, channel modeling, and link performance modeling. The link-performance-modeling block models the \gls{PHY} components of the communication system. Some components of this block (channel encoding, detection, channel decoding, etc.) are time-intensive and computation-intensive to simulate. So, in order to reduce the complexity of \gls{SLS}, these components are replaced by simpler functionalities that are quick to execute but capture the essential behavior of the overall physical layer. This technique is called \gls{PHY} abstraction. The precise goal of \gls{PHY} abstraction is to obtain the same figures of merit for performance evaluation as would be obtained if the original components were used, but with lower complexity. In the literature, there exist abstraction models for \gls{SISO} systems (\cite{lagen2021new}, and references therein), and (with some limitations) for \gls{MU-MIMO} systems with linear receivers. These work by mapping post-equalization \gls{SINR} to \gls{CER}/\gls{BLER}. However, there is no known technique to perform \gls{PHY} abstraction for arbitrary non-linear \gls{MU-MIMO} receivers since there is no known metric equivalent to post-equalization \gls{SINR} for such receivers.     

In this two-part paper, we address the aforementioned issues. The contributions of this part of the two-part paper may be summarized as follows:
\begin{itemize}
\item We introduce the notion of \gls{BMDR} of a \gls{MIMO} detector for a given channel realization or a set of channel realizations. \Gls{BMDR} is our proposed equivalent of post-equalization \gls{SINR} for arbitrary receivers. We relate this \gls{BMDR} to the mismatched decoding frameworks of \cite{Merhav1994, Ganti2000,Martinez2009}, and establish the relationship between \gls{BMDR}, channel code-rate, and \gls{CER} (see Theorem~\ref{thm:main_thm} in Section~\ref{sec:bmdr}).
\item We present a machine-learning-based method to estimate the \gls{BMDR} of a detector for a set of observed channel realizations. The need for a machine-learning-based approach is due to the fact that \gls{BMDR} does not have a closed-form expression or any other straightforward method of calculating it in real time. 
\item We provide simulations results to highlight the efficacy of our proposed method.
\end{itemize} 

In the second part~\cite{kps_jh_part2}, the results obtained here are utilized to describe new techniques to perform \gls{LA} and \gls{PHY} abstraction in \gls{MU-MIMO} systems with arbitrary receivers. 

\subsection{Related Literature}
There have been a few metrics in the literature that can be used in lieu of \gls{SINR} for single user systems. The important ones are \gls{MI}~\cite{Cirkic2011}, capacity of coded-modulation~\cite{Caire1998}, and capacity of \gls{BICM}~\cite{Caire1998}. However, these are general information-theoretic metrics for a chosen input constellation, and do not take into consideration the effect of a specific (possibly suboptimal) receiver. The best known results on information-theoretic achievable rates with mismatched decoding, called \gls{GMI}, can be found in~\cite{Merhav1994, Ganti2000, Martinez2009, Scarlett2014}. However, these results are not directly applicable to the problem in hand since they pertain to single user systems without fading. Other works~\cite{Fertl2008, krishna_sayana} have proposed various information-theoretic metrics for a given \gls{MIMO} detector, but the metrics considered in these papers require the distribution of \glspl{LLR} which is not straightforward to compute in real time. Moreover, the approximation in~\cite{krishna_sayana} holds for a specific $2\times2$ \gls{MIMO} system with an \gls{ML} receiver. To the best of our knowledge, there is no work in the literature that presents a metric equivalent to post-equalization \gls{SINR} for arbitrary \gls{MU-MIMO} receivers in a fading environment, along with a practical method to compute it in real time. 

\subsection*{Paper Organization}
The system model and a few relevant definitions are presented in Section~\ref{sec:system_model}. Section~\ref{sec:bmdr} introduces the \gls{BMDR} and analyses its relevance to error performance. Section~\ref{sec:bmdr_prediction} describes the challenges involved in predicting \gls{BMDR}, while Section~\ref{sec:training} details how machine-learning can be employed to perform \gls{BMDR}-prediction. Simulation results showing the efficacy of the proposed BMDR-prediction method are presented in Section~\ref{sec:sim_results}, and concluding remarks are made in Section~\ref{sec:conc_remarks}.

\subsection*{Notation}
Boldface upper-case (lower-case) letters denote random matrices (vectors), and normal upright upper-case (lower-case) letters are understood from context to denote the realizations of random matrices (vectors). Similarly, if $\Bbc$ denotes a set of random variables, $\Bc$ denotes the set of the realizations of the individual random variables. The field of complex numbers, the field of real numbers, and the ring of integers are respectively denoted by $\Cbb$, $\Rbb$, and $\Zbb$. For any set $\Sc$, $\vert \Sc \vert$ denotes its cardinality if it is finite, and the subset of positive elements in $\Sc$ is denoted by $\Sc_+$. The notation $\Xm \in \Sc^{m \times n}$ denotes that $\Xm$ is a matrix of size $m \times n$ with each entry taking values from a set $\Sc$. For a matrix $\Xm$, its transpose and Hermitian transpose are respectively denoted by $\Xm^{\mathrm{T}}$ and $\Xm^{\mathrm{H}}$, Frobenius norm by $\Vert \Xm \Vert$, and its $(i,j)^{th}$ entry by $[\Xm]_{i,j}$ or $\Xm_{i,j}$ depending on convenience. The block-diagonal matrix with diagonal blocks $\Dm_1,\Dm_2,\cdots,\Dm_n$ is denoted by $\mathrm{diag}\LB \Dm_1,\Dm_2,\cdots,\Dm_n \RB$, and the same notation is used for denoting a diagonal matrix with diagonal elements $d_1,\cdots,d_n$. The identity matrix of size $n\times n$ is denoted by $\Id_n$. The notation $\nv \sim \Cc\Nc( 0, \Id_n)$ and $\nv \sim \Nc( 0, \Id_n)$ respectively denote that $\nv$ is sampled from the $n$-dimensional complex standard normal distribution and the standard normal distribution, while $X \sim \Uc(\Sc)$ denotes that $X$ is a scalar random variable that is sampled uniformly from a set $\Sc$. For a matrix $\Xm \in \Cbb^{m \times n}$ and a vector $\xv \in \Cbb^{n \times 1}$,
\begin{align}\label{eq:notation}
\Xm^{\Rc}  \triangleq  \begin{bmatrix}
\Re(\Xm) & -\Im(\Xm) \\
\Im(\Xm) &  \Re(\Xm)
\end{bmatrix}\in \Rbb^{2m \times 2n}, ~~~ \xv^{\Rc} \triangleq  \begin{bmatrix}
\Re(\xv) \\
\Im(\xv) 
\end{bmatrix} \in \Rbb^{2n \times 1} ,
\end{align}
where $\Re(\Xm) \in \Rbb^{m \times n}$ and $\Im(\Xm) \in \Rbb^{m \times n}$ denote the entry-wise real and imaginary parts of $\Xm$, respectively. The map of an element $x$ in the set $\Xc$ according to a predefined one-to-one mapping rule is denoted by $\mathrm{MAP}\{x ; \Xc\}$, and $x$ can be a tuple. For example, in constellation mapping, a tuple of bits $(b_1,b_2,\cdots,b_m)$ is mapped to a $2^m$-QAM constellation $\Qc$ with the map denoted by  $\textrm{MAP}\{(b_1,b_2,\cdots,b_m);\Qc\}$. The notation $\mathrm{APPEND}\LB \Xc ; x\RB$ denotes the operation of appending an element $x$ to a set $\Xc$. Finally, $\mathrm{SAMPLE}(D)$ denotes a sample drawn from a distribution $D$. The natural logarithm is denoted by $\ln{}$.
\section{System Model and Definitions}
\label{sec:system_model}

\begin{figure}[t]
  \centering
          \includegraphics[height=0.3\textheight, width=\textwidth]{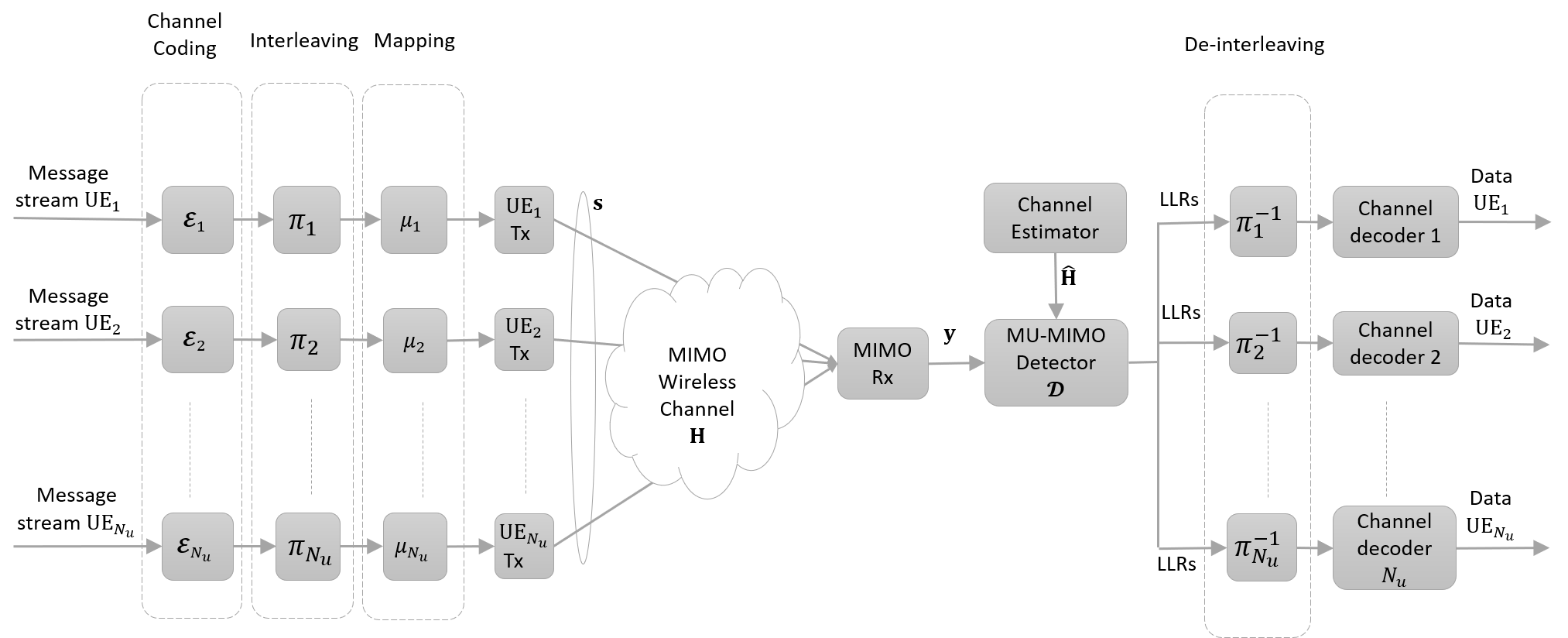}
      \caption{Uplink transmission in a general multi-user MIMO system.}
      \label{fig:mu_mimo_system}
  \end{figure}

We consider an OFDM-based \gls{MIMO} communication system in this paper, but the following technical content is applicable to any communication system in general. Suppose that there are $n_r \geq 1$ receive antennas and $N_u \geq 1$ \glspl{UE}, with \gls{UE} $i$ equipped with $n_t^{(i)}$ transmit antennas, $i = 1,\cdots, N_u$. Let $\sum_{i=1}^{N_u} n_t^{(i)} = N \leq n_r$. A block diagram of the uplink transmission in such a system is depicted in Fig.~\ref{fig:mu_mimo_system}. This setup could also be applied for the downlink.

We consider the classic \gls{BICM}~\cite{Caire1998} for \gls{MU-MIMO} systems. For \gls{UE} $i$, the message set  $\Wc_i = \{w_j, j=1,\cdots, 2^{k_i}\}$ consists of $2^{k_i}$ messages for some positive integer $k_i$.  The channel encoder $\Ec_i$, given as,
\be \label{eq:encoder}
\begin{split}
\Ec_i : \Wc_i  & \longrightarrow \Cc_i \subset \{0,1\}^{n_i}, \\ 
         w & \longmapsto  \Ec_i\LB w \RB \in \Cc_i,
\end{split}
\ee
maps each message to an $n_i$-bit codeword from an $(n_i,k_i)$ code, thereby yielding a code-rate of $k_i/n_i \leq 1$. The codeword bits are then interleaved across the entire codeword using an interleaver denoted by $\pi_i$. The interleaver essentially permutes the order of the codeword bit sequence, and hence, is invertible at the receiver. 

The OFDM system uses $N_{sc}$ subcarriers, and a \gls{RE} consists of one subcarrier and one time symbol. Throughout the rest of the paper, we use the pair $(f,t)$ to index an \gls{RE}, with $f$ denoting the subcarrier index and $t$ denoting the symbol. We assume that \gls{UE} $i$ uses a unit-energy complex constellation for modulation, denoted by $\Qc_i$. The size of $\Qc_i$ is $2^{m_i}$ for some $m_i \in 2\Zbb_+$. We do not consider transmit-precoding in this paper and assume that the number of spatial streams transmitted by each \gls{UE} is equal to the number of its transmit antennas. However, a generalization to the case where there is transmit-precoding and the number of spatial streams is lower is straightforward by integrating the precoder in the channel matrix. Assuming that $m_in_t^{(i)}$ divides $n_i$, every \gls{RE} is associated with $m_in_t^{(i)}$ bits of \gls{UE} $i$, and the $n_i$ interleaved codeword bits are transmitted over $n_i/\LB m_in_t^{(i)}\RB$ \gls{RE}s. Let the interleaved bits of the codeword $\Ec_i\LB w \RB$ that are transmitted on the \gls{RE} indexed by the pair $(f,t)$ be denoted by $b_{f,t,i,l,j}(w)$, or simply $b_{f,t,i,l,j}$, $l=1,\cdots,n_t^{(i)}$, $j=1,\cdots,m_i$. We denote the matrix of these bits by $\Bc_{f,t,i}$ with its $(l,j^{th})$ entry denoted by $b_{f,t,i,l,j}$, $l=1,\cdots,n_t^{(i)}$, $j=1,\cdots,m_i$, and we denote by $\Bbc_{f,t,i}$ the random matrix (of random bits $\boldsymbol{b}_{f,t,i,l,j}$)  whose realizations are $\Bc_{f,t,i}$. It is a common practice to use a {\it scrambler}~\cite[5.3.1]{3GPP_modulation_2020} on the codeword bits in current generation communication systems. Next, a mapper $\mu_i$, which is a bijective map defined as
\be\label{eq:mapper}
\begin{split}
 \mu_i : \{0,1 \}^{n_t^{(i)} \times m_i} & \longrightarrow \Qc_i^{n_t^{(i)} \times 1}, \\ 
  \Bc_{f,t,i} & \longmapsto  \srv_{f,t,i} \in \Qc_i^{n_t^{(i)} \times 1},
\end{split}
\ee
is used to map $\Bc_{f,t,i}$ (or its scrambled bits) to a constellation symbol vector $\srv_{f,t,i} \in \Qc_i^{n_t^{(i)} \times 1}$ by mapping groups of $m_i$ bits to a point in $\Qc_i$. The \gls{UE} then uses a total transmit power $\rho_i$ to transmit this symbol vector over the channel with channel matrix $\Hm_{f,t,i} \in \Cbb^{n_r \times n_t^{(i)}}$. As mentioned earlier, when \gls{UE} $i$ uses a transmit-precoder $\Wm_i \in \Cbb^{n_t^{(i)} \times n_l^{(i)}}$ to transmit $n_l^{(i)} \leq n_t^{(i)}$ spatial streams, its effective channel matrix is $\Hm_{f,t,i}\Wm_i$. We do not assume the availability of \gls{CSI} at the transmitters. So, the power is equally allocated across all the subcarriers. Let $\Gc_i$ denote the set of \gls{RE} index pairs over which the codeword for \gls{UE} $i$ is transmitted. After cyclic-prefix removal and discrete Fourier transform at the receiver, the complex-baseband signal model for an \gls{RE} indexed by $(f,t)$ is given as 

\begin{equation}\label{eq:signal_model}
\yv_{f,t} = \sum_{i=1}^{N_u}\sqrt{\frac{\rho_i}{n_t^{(i)}}}\Hm_{f,t,i}\sv_{f,t,i} + \nv_{f,t} = \Hm_{f,t}\sv_{f,t} + \nv_{f,t} 
\end{equation}
where $\Hm_{f,t} \triangleq \LSB \sqrt{{\rho_1}/{n_t^{(1)}}}\Hm_{f,t,1}, \cdots, \sqrt{{\rho_{N_u}}/{n_t^{(N_u)}}}\Hm_{f,t,N_u} \RSB \in \Cbb^{n_r \times N}$, $\sv_{f,t} \triangleq [\sv_{f,t,1}^{\mathrm{T}}, \cdots, \sv_{f,t,N_u}^{\mathrm{T}} ]^{\mathrm{T}}$ $\in \Qc^{N \times 1}$ with $ \Qc^{N \times 1} \triangleq \Qc_1^{n_t^{(1)} \times 1} \times \cdots \times \Qc_{N_u}^{n_t^{(N_u)} \times 1} $, and $\nv_{f,t} \sim \Cc\Nc\LB 0, \Id_{n_r}\RB$ represents the complex \gls{AWGN}. 

Let $\Ybc_{\Gc_i} = \{\yv_{f,t}, \forall(f,t) \in \Gc_i \}$ and $\Hbc_{\Gc_i} \triangleq \{\Hm_{f,t}, \forall (f,t) \in \Gc_i\}$ respectively be the set of all received signal vectors and the set of channel matrices corresponding to the transmission of an entire codeword of \gls{UE} $i$. We emphasize that $\Ybc_{\Gc_i}$ and $\Hbc_{\Gc_i}$ are random vectors and matrices whose realizations are respectively denoted by $\Yc_{\Gc_i} \triangleq \{\yrv_{f,t}, (f,t) \in \Gc_i \}$ and $\Hc_{\Gc_i} \triangleq \{\Hrm_{f,t}, (f,t) \in \Gc_i \}$. We assume that $\Hc_{\Gc_i}$ is perfectly known at the receiver.

\begin{definition}
(Multi-user MIMO Detector) For the MU-\gls{MIMO} system as described and for the set of constellations $\{\Qc_i\}_{i=1}^{Nu}$, a detector $\Dc$ which is parameterized by $n_r$ and $\{n_t^{(i)}\}_{i=1}^{Nu}$, is a set of $N_u$ maps $\Dc^{(i)}$, $i=1,\cdots,N_u$, defined as follows. 
\be \label{eq:detector}
\begin{split}
\Dc^{(i)} : \Cbb^{n_r\times 1} \times \Cbb^{n_r \times N} \times \Qc^{N \times 1} & \longrightarrow \Rbb^{n_t^{(i)} \times m_i}, \\ 
\LB \yrv_{f,t}, \Hrm_{f,t};\Qc^{N \times 1}\RB & \longmapsto  \Dc^{(i)}\LB \yrv_{f,t}, \Hrm_{f,t};\Qc^{N \times 1} \RB \in \Rbb^{n_t^{(i)} \times m_i}.
\end{split}
\ee
\end{definition}
Henceforth, we won't explicitly state $\Qc^{N \times 1}$, and the constellation set is assumed from context. Here, $\LSB \Dc^{(i)}\LB \yrv_{f,t}, \Hrm_{f,t} \RB \RSB_{l,j}$ is the \gls{LLR} for the $j^{th}$ bit on the $l^{th}$ transmit antenna for \gls{UE} $i$ for the \gls{RE} in context. Let 
\begin{align}\label{eq:qd}
q_{\Dc}(\boldsymbol{b}_{f,t,i,l,j}; \yrv_{f,t}, \Hrm_{f,t}) & \triangleq \frac{1}{1+\exps{-(2\boldsymbol{b}_{f,t,i,l,j}-1)\LSB \Dc^{(i)}\LB \yrv_{f,t}, \Hrm_{f,t} \RB \RSB_{l,j}}}, 
\end{align} 
which is the (possibly mismatched) posterior distribution for $\boldsymbol{b}_{f,t,i,l,j}$ when $\Dc$ is used. Further, let
\begin{align}\label{q_def}
  q^*_{\Dc}(\boldsymbol{b}_{f,t,i,l,j}; \yrv_{f,t}, \Hrm_{f,t}) \triangleq \Prob{\boldsymbol{b}_{f,t,i,l,j} \vert \yv_{f,t}=\yrv_{f,t}, \Hm_{f,t}=\Hrm_{f,t}, \Dc },
\end{align}
which is the true posterior distribution for $\boldsymbol{b}_{f,t,i,l,j}$ when $\Dc$ is used. In general, we have 
\begin{align}\label{eq:LLR_detector}
  \ln{\frac{q^*_{\Dc}(\boldsymbol{b}_{f,t,i,l,j}=1; \yrv_{f,t}, \Hrm_{f,t})}{q^*_{\Dc}(\boldsymbol{b}_{f,t,i,l,j}=0; \yrv_{f,t}, \Hrm_{f,t})}} = \LSB \Dc^{(i)}\LB \yrv_{f,t}, \Hrm_{f,t} \RB \RSB_{l,j} + \varepsilon\LB \Dc  \RB
\end{align}
where $\varepsilon\LB \Dc \RB$ is an error term dependent on $\Dc$. Note that since $\Dc$ is an arbitrary detector and possibly suboptimal, $q^*_{\Dc}(\boldsymbol{b}_{f,t,i,l,j}; \yrv_{f,t}, \Hrm_{f,t})$ itself is a possibly mismatched posterior distribution with respect to the true posterior distribution obtained by an optimal (maximum-likelihood) detector. Additionally, $\varepsilon\LB \Dc  \RB$ accounts for a further mismatch in detection. In many cases, it may be possible to correct this second source of mismatch through deterministic scalar mappings or lookup tables that act on $\Dc^{(i)}\LB \yrv_{f,t}, \Hrm_{f,t} \RB$ \cite{Dijk2003,Jalden2010}. So, in such cases, it may be possible to estimate $q^*_{\Dc}(\boldsymbol{b}_{f,t,i,l,j}; \yrv_{f,t}, \Hrm_{f,t})$ with reasonable accuracy. Throughout the rest of the paper, $q_{\Dc}(\boldsymbol{b}_{f,t,i,l,j}; \yrv_{f,t}, \Hrm_{f,t})$ corresponds to the uncorrected \glspl{LLR} and $q^*_{\Dc}(\boldsymbol{b}_{f,t,i,l,j}; \yrv_{f,t}, \Hrm_{f,t})$ to the corrected \glspl{LLR} for $\Dc$.

Finally, the LLRs for all the codeword bits, after de-interleaving and de-scrambling, are used as inputs to the channel decoder which reconstructs the transmitted message stream. This procedure is called {\it bit-metric decoding}~\cite{bocherer2018achievable} since the \glspl{LLR} are metrics associated with each transmitted bit and are used for decoding. This set-up is in line with that of \cite{Caire1998, Martinez2009}.

\subsection{Imperfect Channel Estimates and Colored Noise}\label{subsec:impf_ch_est}
In practice, one does not have perfect knowledge of the actual channel realizations. Let $\hat{\Hm}_{f,t}$ denote the estimated channel so that $\Hm_{f,t} = \hat{\Hm}_{f,t} + \Delta{\Hm}_{f,t}$, where $\Delta{\Hm}_{f,t}$ denotes the estimation error. The signal model of~\eqref{eq:signal_model} in the presence of interference noise can be written as
\begin{align}\label{eq:imperf_est1}
\yv_{f,t} = \hat{\Hm}_{f,t}\sv_{f,t} + \Delta{\Hm}_{f,t}\sv_{f,t} + \nv_{f,t}.
\end{align} 
In \eqref{eq:imperf_est1}, the interference noise is subsumed in $\nv_{f,t}$ so that $\nv_{f,t} \sim \Cc\Nc\LB 0, \Km_n \RB$ with $\Km_n \in \Cbb^{n_r \times n_r}$ being a Hermitian, positive-definite but non-diagonal matrix. If an \gls{LMMSE} estimator is used for channel estimation, then, $\Delta{\Hm}_{f,t}$ is uncorrelated with $\Hm_{f,t}$~\cite{Kay1993} from the orthogonality principle. Further, if one has an estimate of the mean and the correlation matrix of $\Hm_{f,t}$, it is straightforward to estimate the error correlation matrix $\Km_e \triangleq \EE\LP \Delta{\Hm}_{f,t}\Delta{\Hm}_{f,t}^{\mathrm{H}}\RP \in \Cbb^{n_r \times n_r}$. This can also be based on the instantaneous channel statistics if we assume that $\Hm_{f,t}$ has a non-stationary distribution (see, for example, \cite{goutay2020machine} and references therein). Assuming that $\Hm_{f,t}$ has zero mean, $\Delta{\Hm}_{f,t}\sv_{f,t}$ can be approximated by a zero-mean Gaussian noise vector with covariance $\Km_e$ since the constellations $\Qc_i$ are all assumed to be of unit energy. Noting that $\Km_n + \Km_e $ is positive-definite and Hermitian, its eigendecomposition yields $\Km_n + \Km_e = \Um \mathbf{\Lambda}\Um^{\mathrm{H}}$ with $\mathbf{\Lambda}$ being a diagonal matrix with positive elements. Let $\mathbf{\Lambda}^{1/2}$ denote the diagonal matrix with its diagonal entries being the square roots of those of $\mathbf{\Lambda}$. Therefore, after noise-whitening, the signal model can be expressed as 
\begin{align}\label{eq:imperf_est2}
\yv'_{f,t} = \Hm'_{f,t}\sv_{f,t} + \nv_{f,t}' 
\end{align} 
where $\yv'_{f,t} = \LB \Km_n + \Km_e \RB^{-\frac{1}{2}}\yv_{f,t}$, $\Hm'_{f,t} = \LB \Km_n + \Km_e \RB^{-\frac{1}{2}}\hat{\Hm}_{f,t}$, and $\nv'_{f,t} \sim \Cc\Nc\LB 0, \Id_{n_r}\RB$ due to noise-whitening. Here, $\LB \Km_n + \Km_e \RB^{-\frac{1}{2}} =  \Um \mathbf{\Lambda}^{-\frac{1}{2}}\Um^{\mathrm{H}}$. The equivalent signal model in~\eqref{eq:imperf_est2} is similar to the one in~\eqref{eq:signal_model} with the difference being a deterioration in the \gls{SINR} for each user.

In general, we denote by $q_{\Dc, CE}(\boldsymbol{b}_{f,t,i,l,j}; \yrv'_{f,t}, \Hrm'_{f,t})$ the posterior distribution of bit $\boldsymbol{b}_{f,t,i,l,j}$ when detector $\Dc$ is used for a given channel estimation technique $CE$ that outputs a post-processed channel estimate ${\Hrm'}_{f,t}$ and a post-processed observed signal $\yrv'_{f,t}$, as shown in \eqref{eq:imperf_est2}. Unless otherwise specified, $\Hrm'_{f,t}$ denotes an imperfect post-processed channel estimate of $\Hrm_{f,t}$.

\subsection{\gls{BICM} for \gls{MU-MIMO} with Mismatched Decoding}

The system under consideration is a \gls{BICM} system \cite{Caire1998} with possibly mismatched decoding \cite{Martinez2009,Ganti2000,Merhav1994} when applied to \gls{MU-MIMO} transmission. With $\Bc_{i} \in \{0,1 \}^{n_t^{(i)} \times m_i}$ denoting the matrix of bits to be transmitted by UE $i$ on a \gls{RE}, and the constellation mapper $\mu_i$ as defined in \eqref{eq:mapper} for each $i=1,\cdots, N_u$, let 
\begin{align}
  \Sc^{+}_{i,l,j} \triangleq \LP \sv = [\sv_{1}^{\mathrm{T}}, \cdots, \sv_{N_u}^{\mathrm{T}} ]^{\mathrm{T}}\in \Qc^{N \times 1} \middle| \begin{matrix}
    \sv_{i} = \mu_i\LB\Bc_{i}\RB,   \\
    \forall \Bc_{i} \in \{0,1 \}^{n_t^{(i)} \times m_i} \textrm{ with } b_{i,l,j} = 1 
   \end{matrix} \RP.
\end{align}
Here, $\Sc^{+}_{i,l,j}$ is the set of all composite input signal vectors corresponding to the $(l,j)^{th}$ bit $b_{i,l,j}$ of UE $i$ being $1$. Similarly, $\Sc^{-}_{i,l,j}$ can be defined to be set of all composite input signal vectors corresponding to the $(l,j)^{th}$ bit of UE $i$ being $0$. The \gls{ML}-detector \cite{Studer2010}, denoted by $\Dc_{MLD}$, generates the \gls{LLR} for $b_{f,t,i,l,j}$ as given in \eqref{eq:LLR_detector} as 
\begin{align} \label{eq:MLD}
  \LSB \Dc_{MLD}^{(i)}\LB \yrv_{f,t}, \Hrm_{f,t} \RB \RSB_{l,j} = \ln{\frac{\sum_{\srv_{f,t} \in \Sc^{+}_{i,l,j}}\Prob{\yrv_{f,t}\vert \srv_{f,t}, \Hrm_{f,t}}}{\sum_{\srv_{f,t} \in \Sc^{-}_{i,l,j}}\Prob{\yrv_{f,t}\vert \srv_{f,t}, \Hrm_{f,t}}}}.
\end{align}
When $\Dc_{MLD}$ is used for detection, the system is said to be a \gls{BICM} system with matched decoding \cite{Martinez2009}. However, in a \gls{MU-MIMO} setting, the complexity of $\Dc_{MLD}$ is of the order of $2^{\sum_{i=1}^{N_u}m_in_t^{(i)}}$ which can be practically prohibitive. Therefore, suboptimal detection techniques which include, except under special cases, all linear detectors and the fixed complexity variants of the sphere detector, are employed in practice. In such a case, the \gls{BICM} system is said to use {\it mismatched} decoding. 

Consider an arbitrary bit decoding metric $q_{\Dc}(\boldsymbol{b}_{f,t,i,l,j}; \yrv_{f,t}, \Hrm_{f,t})$ of the form shown in \eqref{eq:qd}. For the set-up under consideration in the case of static channels, i.e., $\Hm_{f,t} = \Hrm_{f}$ which is invariant in time, the largest achievable transmission rate $k_i/n_i$ of the encoding scheme given by \eqref{eq:encoder} for $q_{\Dc}$ which guarantees a \gls{CER} arbitrarily close to $0$ as $n_i \to 0$ is lower-bounded by the \gls{GMI} \cite{Martinez2009,Merhav1994, Ganti2000, Lapidoth2002}. This \gls{GMI} is given as 
\begin{align}\label{eq:GMI}
I_{gmi} \triangleq \sup_{s>0}I_{gmi}(s)
\end{align}
where $I_{gmi}(s)$ in the \gls{MU-MIMO} set-up can be derived to be 
\begin{align}\label{eq:gmi_s}
  I_{gmi}(s) \triangleq \max\LP \frac{1}{n_i}\sum_{(f,t)\in \Gc_i}\Expect{\yv_{f,t}}{\sum_{l=1}^{n_t^{(i)}} \sum_{j=1}^{m_i}\log{\frac{q_{\Dc}\LB \boldsymbol{b}_{f,t,i,l,j}; \yv_{f,t}, \Hrm_{f} \RB^s}{\frac{1}{2}\sum_{b=0}^1 q_{\Dc}\LB b; \yv_{f,t}, \Hrm_{f}\RB^s}}}, 0 \RP.
\end{align}
In \eqref{eq:gmi_s}, $\LP \boldsymbol{b}_{f,t,i,l,j} \RP_{l,j}$ is related to the elements of $\sv_{f,t,i}$ through the bijective map $\mu_i$, and the expectation is over $\yv_{f,t}$ which is dependent only on $\sv_{f,t}$ and $\nv_{f,t}$ for a fixed $\Hrm_{f}$. A decoder of the form 
\begin{align}\label{eq:bit_metric_decoding}
  \hat{w} = \argmax{w \in \Wc_i}\prod_{(f,t)\in \Gc_i}\prod_{l=1}^{n_t^{(i)}} \prod_{j=1}^{m_i} q_{\Dc}(b_{f,t,i,l,j}(w); \yrv_{f,t}, \Hrm_{f})
\end{align}
is shown to achieve $I_{gmi}$ \cite{Martinez2009}.
\section{Bit-Metric Decoding Rate}
\label{sec:bmdr}

In this section, we introduce the notion of \gls{BMDR} of a detector for a user when paired with other users. The \gls{BMDR} of a detector $\Dc$ for \gls{UE} $i$ for a channel matrix $\Hm_{f,t}$, denoted by $R_{\Dc,i}(\Hm_{f,t})$, is defined as follows:
\begin{eqnarray} \label{eq:gen_bmdr_channel_def} 
    R_{\Dc,i}(\Hm_{f,t})  & \triangleq  & \max\LP 1 +\Expect{\yv_{f,t} \vert \Hm_{f,t} }{\frac{1}{m_in_t^{(i)}}\sum_{l=1}^{n_t^{(i)}}\sum_{j=1}^{m_i} \log{q_{\Dc}(\boldsymbol{b}_{f,t,i,l,j}; \yv_{f,t}, \Hm_{f,t})}}, 0\RP
   \end{eqnarray}
where $\LP \boldsymbol{b}_{f,t,i,l,j} \RP_{l,j}$ is related to the elements of $\sv_{f,t,i}$ through the bijective map $\mu_i$, and $\yv_{f,t}$ is dependent only on $\sv_{f,t}$ and $\nv_{f,t}$ when conditioned on $\Hm_{f,t}$. Note that $R_{\Dc,i}(\Hm_{f,t})$ is itself a random variable whose realization is dependent on the realization of the random channel matrix $\Hm_{f,t}$. However, conditioned on a channel realization, $R_{\Dc,i}(\Hm_{f,t})$ can be obtained by plugging in $s=1$ in \eqref{eq:gmi_s}. With \gls{LLR} correction so that $\varepsilon\LB \Dc  \RB$ in \eqref{eq:LLR_detector} is corrected for,
\begin{eqnarray} \label{eq:cor_bmdr_channel_def} 
    R^*_{\Dc,i}(\Hm_{f,t})  & \triangleq  & \max\LP 1 +\Expect{\yv_{f,t} \vert \Hm_{f,t} }{\frac{1}{m_in_t^{(i)}}\sum_{l=1}^{n_t^{(i)}}\sum_{j=1}^{m_i} \log{q^*_{\Dc}(\boldsymbol{b}_{f,t,i,l,j}; \yv_{f,t}, \Hm_{f,t})}}, 0\RP.
   \end{eqnarray}
It follows that for any arbitrary $\Dc$,  $R^*_{\Dc,i}(\Hm_{f,t}) \geq R_{\Dc,i}(\Hm_{f,t})$, $\forall \Hm_{f,t}$. This is seen by noting that $R^*_{\Dc,i}(\Hm_{f,t}) - R_{\Dc,i}(\Hm_{f,t}) = D_{KL}\LB q^*_{\Dc} \Vert q_{\Dc} \RB \geq 0$, the Kullback-Leibler divergence between the true posterior $q^*_{\Dc} $ and the possibly mismatched posterior $ q_{\Dc}$ when $\Dc$ is used.

The BMDR of a detector $\Dc$ for a set of channel matrices $\Hbc$ is defined as
\begin{eqnarray} \label{eq:gen_bmdr_def} 
    R_{\Dc,i}(\Hbc)  & \triangleq  &  \frac{1}{\vert \Hbc \vert}\sum_{\Hm_{f,t} \in \Hbc}  R_{\Dc,i}(\Hm_{f,t}),
   \end{eqnarray} 
and likewise for the \gls{LLR}-corrected metrics $q^*_{\Dc}$. When a practical channel estimation method $CE$ is employed so that only an imperfect, post-processed channel estimate $\Hm'_{f,t}$ of $\Hm_{f,t}$ is available along with the post-processed observed signal $\yv'_{f,t}$, the expression for $R_{\Dc,i}(\Hm_{f,t})$ in \eqref{eq:gen_bmdr_channel_def} changes only slightly by replacing $q_{\Dc}(\boldsymbol{b}_{f,t,i,l,j}; \yv_{f,t}, \Hm_{f,t})$ with $q_{\Dc, CE}(\boldsymbol{b}_{f,t,i,l,j}; \yv'_{f,t}, \Hm'_{f,t})$, and we denote the resulting \gls{BMDR} by $R_{\Dc,CE,i}(\Hm_{f,t})$. 

Note that $R_{\Dc,i}(\Hm_{f,t})$ is a monotone non-decreasing function of the transmit power $\rho_i$ of each user because the posterior distribution of a bit can only improve with an increase in the distance between constellation points. For linear detectors, post-equalization \gls{SINR} is monotone increasing with the transmit power. {\it Therefore, \gls{BMDR} is monotone non-decreasing with post-equalization \gls{SINR} for linear detectors.}

Let $\Gc_i(n)$ denote a set of \gls{RE} index pairs of cardinality $n$, and $\Hc_{\Gc_i(n)}$ the associated set of channel realizations over which a codeword of \gls{UE} $i$ is transmitted. We are interested in the codeword error behavior of \gls{UE} $i$ when detector $\Dc$ is used to generate the \glspl{LLR}.

\begin{thm} \label{thm:main_thm}
Consider a sequence of sets of channel realizations $\Hc_{\Gc_i(n)} = \LP \Hrm_1,\cdots \Hrm_n \RP $, $n\in \Zbb_+$, such that $0 < R_{\Dc,i}^{min} \triangleq \inf\limits_{n \in \Zbb_+} R_{\Dc,i}(\Hc_{\Gc_i(n)}) \leq \sup\limits_{n \in \Zbb_+} R^*_{\Dc,i}(\Hc_{\Gc_i(n)}) = R_{\Dc,i}^{max^*}$ for a detector $\Dc$ defined by \eqref{eq:detector}. Suppose that the codeword bits of \gls{UE} $i$ of blocklength $n_i$ are to be transmitted over $\Hc_{\Gc_i(n)}$ with $n_i = nm_in_t^{(i)}$. Then, there exists an $(n_i, k_i)$ code for which the probability of codeword error is arbitrarily close to $0$ as $n_i \to \infty$ if $\frac{k_i}{n_i} < R_{\Dc,i}^{min}$. Further, for any code with $\frac{k_i}{n_i} > R_{\Dc,i}^{max^*}$, the \gls{CER} is bounded away from $0$.
\end{thm}

\begin{figure}
\centering
        \includegraphics[scale = 0.53]{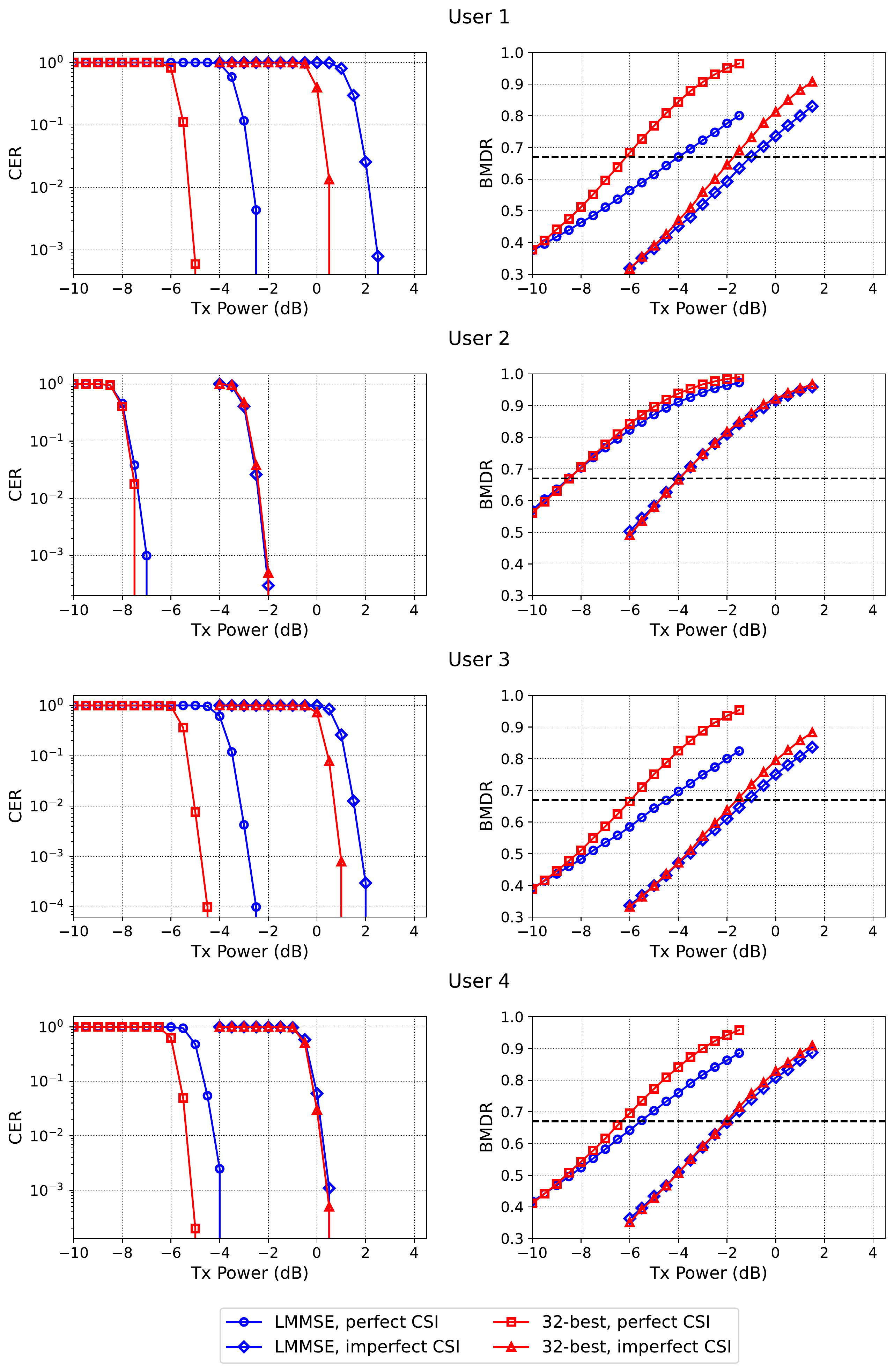}
    \caption{The \gls{CER} (left) and \gls{BMDR} (right) comparison for a fixed set of channel realizations with both perfect \gls{CSI} and with imperfect (\gls{LMMSE}-estimated) \gls{CSI}.}
    \label{fig:bmdr_cer_certain_channels}
\end{figure}

The proof is provided in the Appendix. {\it The main message of the theorem is that if a codeword is to be transmitted over a finite set of channel realizations $\Hc_{\Gc_i}$ with \gls{BMDR} $R_{\Dc,i}(\Hc_{\Gc_i})$, a code-rate of $R_{\Dc,i}(\Hc_{\Gc_i})$ is achievable while the maximum code-rate for reliable transmission is upper-bounded by $R^*_{\Dc,i}(\Hc_{\Gc_i})$ which itself may be achieved by \gls{LLR} correction.} 

We can interpret $R_{\Dc,i}(\Hc_{\Gc_i})$ as follows: If the set of channel realizations $\Hc_{\Gc_i}$ were fixed for several codeword transmissions and a detector $\Dc$ is used, then, \gls{UE} $i$ could transmit reliably using a code-rate at most equal to $R_{\Dc,i}(\Hc_{\Gc_i})$. To illustrate this, we consider a simulation setup using the QuaDRiGa channel-simulator~\cite{Jaeckel2014}. A $(1800,1200)$ 5G \gls{LDPC} code is employed along with QPSK modulation ($4-$QAM). The base station is set up to have $16$ antennas, and there are four users placed at random within a distance of \SI[input-quotient  = :,output-quotient = \text{--}]{25:250}{\metre} from the base station. The exact details of the channel-generator scenario are presented in Section~\ref{sec:sim_results}. A sequence of $900$ \gls{MU-MIMO} channel matrices is generated according to the urban macro-cell (UMa) scenario with non-line-of-sight (NLoS) propagation. Let $\{\hrv_{k}^{(i)} \in \Cbb^{16 \times 1}, k=1,\cdots,900 \}$ denote the channel vectors from \gls{UE} $i$ to the base station, $i=1,2,3,4$. We also have

\begin{align}
\frac{1}{900}\sum_{k=1}^{900} \left\Vert \hrv_{k}^{(i)} \right\Vert^2 = 16, ~~~i=1,2,3,4.
\end{align}	 

For each user, $10,000$ independent codewords are transmitted over the user's sequence of channel vectors. In particular, each codeword of a user sees the same set of $900$ channel vectors but different noise realizations. This is of course not representative of a realistic wireless fading environment where the channel matrices are different for different codeword transmissions, but considered here only to support the claims of Theorem~\ref{thm:main_thm}. Assuming that all the users use the same transmit power $\rho$, the \gls{CER} as a function of $\rho$ is plotted in Fig.~\ref{fig:bmdr_cer_certain_channels} for each of the four users with the LMMSE detector and the $32$-best detector employed ($K=32$ for the $K$-best detector~\cite{Guo2006}). The plots are for both perfect \gls{CSI} at the receiver with the signal model given by~\eqref{eq:signal_model}, and with \gls{LMMSE}-estimated \gls{CSI} at the receiver with the equivalent signal model given by~\eqref{eq:imperf_est2}. 

Next, we empirically estimate the \gls{BMDR} for the two detectors as a function of $\rho$. The methodology to estimate \gls{BMDR} empirically is described in Section~\ref{sec:bmdr_prediction}, and the \gls{BMDR} approximations for each composite channel are given by~\eqref{eq:mc_approx} and \eqref{eq:mc_approx_CE} for perfect CSI and imperfect CSI, respectively. Let $\Hrm_{k}(\rho) \triangleq \rho \LSB \hrv_{k}^{(1)},\cdots, \hrv_{k}^{(4)} \RSB \in \Cbb^{16\times 4}$ and $\Hc(\rho) \triangleq \LP  \Hrm_{k}(\rho), k=1,\cdots,900 \RP$. We estimate
\begin{align}
    \bar{R}_{\Dc,i}\LB \Hc(\rho) \RB = \frac{1}{900}\sum_{k=1}^{900}\bar{R}_{\Dc,i} \LB \Hrm_{k}(\rho) \RB, ~~\forall i=1,\cdots,4,
\end{align}
using \eqref{eq:mc_approx}, and 
\begin{align}
    \bar{R}_{\Dc,LMMSE,i}\LB \Hc(\rho) \RB = \frac{1}{900}\sum_{k=1}^{900}\bar{R}_{\Dc,LMMSE, i} \LB \Hrm_{k}(\rho) \RB,~~ \forall i=1,\cdots,4,
\end{align} 
using \eqref{eq:mc_approx_CE}, with $N_{samp}=1000$. Here,  $\bar{R}_{\Dc,LMMSE, i}(\Hrm_{k}(\rho))$ denotes the $i^{th}$ user's \gls{BMDR} with \gls{LMMSE} channel estimation, and $\Dc$ is the \gls{LMMSE}/$32$-best detector as the case may be. Plotted to the right of each \gls{CER} plot for each user is the \gls{BMDR} for the two detectors as a function of $\rho$. It can clearly be seen that for all the \gls{CER} curves, the fall begins only after a transmit power corresponding to a \gls{BMDR} of $0.67$ (indicated by the horizontal dashed line in the BMDR plots) has either been reached or exceeded. This is true for the case of perfect \gls{CSI} as well as for the case of imperfect \gls{CSI}. Since we are using a rate-$2/3$ \gls{LDPC} code, a fall in \gls{CER} can occur only when the \gls{BMDR} exceeds $0.67$, as a direct consequence of Theorem~\ref{thm:main_thm}.

\section{BMDR prediction}
\label{sec:bmdr_prediction}
The result of Theorem~\ref{thm:main_thm} offers some insights into the error-probability behavior of a particular detector for a given set of channel realizations. {\it To be precise, $R_{\Dc,i}(\Hc_{\Gc_i})$ is an achievable rate only if the channel is static}. Nevertheless, the channel does not change drastically within a short instant of time for slow-moving users in an urban environment. Therefore, if one has previous channel estimates of the users, the result of Theorem \ref{thm:main_thm} can be exploited for important tasks like link adaptation in a communication system. For any such task, it is vital to predict $R_{\Dc,i}(\Hc_{\Gc_i})$ for a set of given channels $\Hc_{\Gc_i}$ or their estimates. Likewise, if \gls{LLR} correction is applied, the goal is to predict $R^*_{\Dc,i}(\Hc_{\Gc_i})$. In the rest of the paper, we assume that there is no \gls{LLR} correction, but the following technique applies to the latter case as well.

Unfortunately, the exact expression for $R_{\Dc,i}(\Hrm_{f,t})$ as given by~\eqref{eq:gen_bmdr_channel_def} is not computable even for the simplest of linear detectors even though closed form expressions for $q_{\Dc}$ exist. However, $R_{\Dc,i}(\Hrm_{f,t})$ can be empirically approximated as follows. Let $\sv  \triangleq [\sv_1^\mathrm{T}, \sv_2^\mathrm{T},\cdots, \sv_{N_u}^\mathrm{T}]^\mathrm{T} \in \Qc^{N \times 1}$ with $\sv_i \in \Qc_i^{n_t^{(i)} \times 1}$ being complex random signal vectors whose realizations are the transmitted constellation signals chosen independently and uniformly from the constellation $\Qc_i^{n_t^{(i)} \times 1}$, $i=1,\cdots,N_u$. It goes without saying that the bits $\{b_{f,t,i,l,j}\}$ also take values according to $\sv_i$. Further, let $\nv \in \Cbb^{n_r \times 1} $ have the standard complex normal distribution. Therefore, by drawing $N_{samp}$ independent samples of $\nv$ and $\sv$, denoted respectively by $\{\nrv^{(r)}, r=1,\cdots, N_{samp}\}$ and $\{\srv^{(r)}, r=1,\cdots, N_{samp}\}$, from their respective distributions, a Monte-Carlo approximation for~\eqref{eq:gen_bmdr_def} is given as
\begin{equation}
\label{eq:mc_approx}
\bar{R}_{\Dc,i}(\Hrm_{f,t}) = \max \LP 1 + \frac{1}{m_i n_t^{(i)}N_{samp}}\sum_{r=1}^{N_{samp}}\LP \sum_{l=1}^{n_t^{(i)}}\sum_{j=1}^{m_i}\log{q_{\Dc}(b^{(r)}_{f,t,i,l,j}; \yrv^{(r)}, \Hrm_{f,t})}\RP, 0\RP
\end{equation} 
where $ \yrv^{(r)} = \Hrm_{f,t}\srv^{(r)} + \nrv^{(r)}$. $r = 1,\cdots, N_{samp}$. When a practical channel estimation technique $CE$ is used that outputs a post-processed channel estimate $\Hrm_{f,t}'$ of $\Hrm_{f,t}$ and a post-processed received data signal $\yrv^{(r)'}$ of $\yrv^{(r)}$, the Monte-Carlo approximation is given as 
\begin{equation}
    \label{eq:mc_approx_CE}
    \bar{R}_{\Dc,CE,i}(\Hrm_{f,t}) = \max \LP 1 + \frac{1}{m_i n_t^{(i)}N_{samp}}\sum_{r=1}^{N_{samp}}\LP \sum_{l=1}^{n_t^{(i)}}\sum_{j=1}^{m_i}\log{q_{\Dc}(b^{(r)}_{f,t,i,l,j}; \yrv^{(r)'}, \Hrm_{f,t}')}\RP, 0 \RP.
\end{equation} 
From Hoeffding's inequality~\cite[Ch. 2]{McDiarmid1998}, it follows that for all $\Hrm_{f,t} \in \Cbb^{n_r \times N}$,
\begin{equation}
\Prob{\left\vert \bar{R}_{\Dc,i}(\Hrm_{f,t}) -  R_{\Dc,i}(\Hrm_{f,t})\right\vert \geq \delta} \leq 2\exp{-\frac{N_{samp}^2\delta^2}{K}}
\end{equation}
for any $\delta > 0$ with $K$ being a positive constant. Now, it would be of prohibitive complexity to perform a Monte-Carlo sampling to obtain the approximation in~\eqref{eq:mc_approx} for every observed channel in real time. Instead, the idea is to train a neural network that takes as input the channel realization $\Hrm_{f,t}$ or some suitable function of $\Hrm_{f,t}$, and outputs an estimate of $R_{\Dc,i}(\Hrm_{f,t})$. Let $f_{\Theta, \Dc, i}$ denote this neural network, with the set of trainable parameters denoted by $\Theta$. Such a network would have to be trained on a dataset consisting of the labeled pairs $\LB g(\Hm), \bar{R}_{\Dc,i}(\Hm) \RB$, $\Hm \in \Cbb^{n_r \times N}$, where $g$ is some suitable function. Since the actual target label for a matrix $\Hrm_{f,t}$ during training is $\bar{R}_{\Dc,i}(\Hrm_{f,t})$, we assume the following model for $f_{\Theta, \Dc, i}$: 
\begin{align}\label{eq:g_function}
f_{\Theta, \Dc, i}(g(\Hm)) = \bar{R}_{\Dc,i}(\Hm) + \varepsilon, ~~\Hm \in \Cbb^{n_r \times N}
\end{align}
where $\varepsilon$ is zero-mean bounded noise. 

\begin{remark}\label{rem:lmmse_pred}
    Without using a machine-learning model, it is possible to estimate the \gls{BMDR} of a linear detector for each user based on the post-equalization \gls{SINR} due to the fact that \gls{BMDR} is a monotone non-decreasing function of post-equalization \gls{SINR}. Such an equalization effectively creates $N$ parallel channels for the $N$ transmitted symbols. By pre-computing (empirically) the \gls{BMDR} for a \gls{SISO} \gls{AWGN} channel with maximum-likelihood decoding for every used constellation over a certain range of \glspl{SNR}, one can obtain a \gls{SNR}-\gls{BMDR} map. Then, the post-equalization \gls{SINR} for each transmitted symbol in the \gls{MIMO} setting is computed and mapped to a \gls{BMDR} value using this pre-computed \gls{SNR}-\gls{BMDR} map. Such an approach works very well for the LMMSE detector. So, a machine-learning model for BMDR prediction is only required for non-linear MU-MIMO detectors.  
\end{remark}

\section{Training a Convolutional Neural Network for BMDR prediction}
\label{sec:training}

In this section, we describe a method to train a \gls{CNN}~\cite{LeCun1999} to predict $R_{\Dc,i}(\Hrm)$ for a given channel matrix $\Hrm \in \Cbb^{n_r \times N}$. Rewriting~\eqref{eq:signal_model} and dropping the \gls{RE} indices, we have
\begin{equation}
\label{eq:signal_model_1}
\yv = \Hm\sv + \nv = \sum_{i=1}^{N_u}\sqrt{\frac{\rho_i}{n_t^{(i)}}}\Hm_i  \sv_i + \nv
\end{equation}
where $\Hm_i \in \Cbb^{n_r \times n_t^{(i)}}$ has $\Expect{\Hm_i}{\Vert \Hm_i \Vert^2} = n_rn_t^{(i)}$, $\Expect{\sv_i}{\sv_i \sv_i^{\mathrm{H}}} = \Id_{n_t^{(i)}}$, $\nv \sim \Cc\Nc\LB 0, \Id_{n_r}\RB$, and $\rho_i$ denotes the transmit power of \gls{UE} $i$. Equivalently, we have the following model in the real domain from the definition in~\eqref{eq:notation}:
\begin{equation}
\yv^{\Rc} = \Hm^{\Rc} \sv^{\Rc} + \nv^{\Rc},
\end{equation}
with $\yv^{\Rc} \in \Rbb^{2n_r \times 1}$, $\nv^{\Rc} \sim \Nc(0, 0.5 \Id_{2n_r})$,  $\sv^{\Rc} \in \Rbb^{2N \times 1}$, and $\Hm^{\Rc} \in \Rbb^{2n_r \times 2N}$. Upon the $\mathbf{QR}$-decomposition of $\Hm^{\Rc}$, we obtain $\Hm^{\Rc} = \Qm\Rm$ where $\Qm \in \Rbb^{2n_r \times 2N}$ is column orthogonal, and $\Rm \in \Rbb^{2N \times 2N}$ is upper-triangular. A linear post-processing performed by left-multiplying $\yv^{\Rc}$ by $ \Qm^{\mathrm{T}}$ results in 
\begin{equation}
\label{eq:eqv_signal_model}
\bar{\yv} \triangleq \Qm^{\mathrm{T}}\yv^{\Rc} = \Rm \sv^{\Rc} + \bar{\nv}
\end{equation}
where $\bar{\yv} \in \Rbb^{2N \times 1}$, and $\bar{\nv} \triangleq \Qm^{\mathrm{T}}\nv^{\Rc} \sim \Nc(0, 0.5 \Id_{2N})$. Clearly, $R_{\Dc,i}(\Hm) = R_{\Dc,i}(\Rm)$ for any detector $\Dc$ and \gls{UE} $i$, where $R_{\Dc,i}(\Rm)$ is computed in the real-domain. But since $N \leq n_r$, $\Rm$ has less non-zero entries than $\Hm$, and also captures all the useful information pertaining to the \gls{BMDR} in the following manner:
\begin{enumerate}
\item The upper off-diagonal elements capture the amount of \gls{ISI} between users' symbols in the same channel use. The stronger their magnitudes relative to the diagonal elements, the smaller the \gls{BMDR}.
\item The magnitudes of all the non-zero elements capture the \gls{SINR} which influences the \gls{BMDR}.
\end{enumerate}  

\begin{algorithm}
\caption{Pseudocode to generate labeled data for a detector $\Dc$ in a \gls{MU-MIMO} system with $n_r$ Rx antennas, $N_u$ users; \gls{UE} $i$ has $n_t^{(i)}$ Tx antennas and uses a unit-energy constellation $\Qc_i$ with $\vert \Qc_i\vert=2^{m_i}$. The minimum and maximum transmit power for \gls{UE} $i$ are $\rho_{min,i}$ and $\rho_{max,i}$ \si{dB}, respectively. Each channel matrix in the dataset generates $N_p$ input feature-label pairs.}
\label{alg:generating_labels}
\begin{algorithmic}
  \Input
  \Desc{$\Hc$}{: Set of composite \gls{MU-MIMO} channel matrices of size $n_r \times N$, $N = \sum_{i=1}^{N_u}n_t^{(i)}$}  
  \EndInput
  \Output
  \Desc{$\Xc_{L, \Dc}$}{: Set of input features}  
  \Desc{$\Yc_{L, \Dc}^{(i)}$}{: Set of target labels for \gls{UE} $i$ with a one-to-one mapping to $\Xc_{L, \Dc}$} 
  \EndOutput 
\Initialize{\strut$\Xc_{L, \Dc} \gets \{\}$, $\Yc_{L, \Dc}^{(i)} \gets \{\}$, $\forall i=1,\cdots,N_u$}   
\ForAll{$\Hrm = [\Hrm_1,\cdots,\Hrm_{N_u}] \in \Hc$}
   \State $\Hrm_{norm,i} \gets \sqrt{n_rn_t^{(i)}}\Hrm_i/\Vert \Hrm_i \Vert$, $\forall i=1,\cdots,N_u$, $\Hrm_{norm} \gets \LSB \Hrm_{norm,1},\cdots,\Hrm_{norm,N_u}\RSB $    
   \State $\mathrm{Q}, \mathrm{R} \gets \mathrm{QR}(\Hrm_{norm}^{\Rc})$ \Comment{QR-decomposition}  
   \For{$k \gets 1$ to $N_{p}$}
      \State $\rho_{i,dB} \gets \mathrm{SAMPLE}\LB\Uc \LB \LSB\rho_{min,i},\rho_{max,i}\RSB \RB\RB$, $\rho_{i} \gets 10^{\rho_{i,dB}/10}$,  $\forall i=1,\cdots, N_u$
      \State $r_{\Dc,i} \gets 0$, $\forall i=1,\cdots,N_u$
      \For{$t \gets 1$ to $N_{samp}$}
         \State $\nrv \gets \mathrm{SAMPLE}\LB\mathcal{CN}(0,\Id_{n_r})\RB$  
            \For{$i \gets 1$ to $N_u$}
               \State $b_{i,l,j} \gets \mathrm{SAMPLE}\LB\Uc(\{0,1\} )\RB$, $\forall l=1,\cdots, n_t^{(i)}$, $\forall j=1,\cdots,m_i$
               \State $s_{i,l} \gets \mathrm{MAP}\LP (b_{i,l,1},\cdots,b_{i,l,m_i}) ; \Qc_i \RP$, $\forall l=1,\cdots,n_t^{(i)}$
               \State $\srv_i \gets [s_{i,1},\cdots,s_{i,n_t^{(i)}}]^{\mathrm{T}}$  
            \EndFor            
            \State $\yrv \gets \sum_{i=1}^{Nu}\sqrt{\frac{\rho_i}{n_t^{(i)}}}\Hrm_{norm,i}\srv_i + \nrv $, $\Omega \gets \mathrm{diag}\LB \sqrt{\frac{\rho_1}{n_t^{(1)}}}\Id_{n_t^{(1)}}, \cdots,\sqrt{\frac{\rho_{N_u}}{n_t^{(N_u)}}}\Id_{n_t^{(N_u)}} \RB$
            \State { $r_{\Dc,i} \gets  r_{\Dc,i} +  \frac{\sum_{l=1}^{n_t^{(i)}}\sum_{j=1}^{m_i}\log{q_{\Dc,i,l,j}(b_{i,l,j}; \yrv, \Hrm_{norm}\Omega)}}{m_i n_t^{(i)}N_{samp}}$, $\forall i=1,\cdots, N_u$ }          
      \EndFor 
      \State $\bar{R}_{\Dc,i} \gets \max\{0, 1+r_{\Dc,i}\}$, $\forall i=1,\cdots, N_u$ 
      \State $\Xc_{L,\Dc} \gets \mathrm{APPEND}\LB \Xc_{L,\Dc} ; \mathrm{R}\Omega^{\Rc} \RB$, $\Yc_{L, \Dc}^{(i)} \gets \mathrm{APPEND}\LB \Yc_{L, \Dc}^{(i)} ;\bar{R}_{\Dc,i} \RB, \forall i=1,\cdots,N_u $ 
   \EndFor
\EndFor
\end{algorithmic}
\end{algorithm}

In essence, $R_{\Dc,i}(\Hm)$ is completely determined by the upper-triangular matrix $\Rm$ when the other parameters $n_r$, $\{n_t^{(i)}, \Qc_i \}_{i=1}^{N_u}$ are fixed. Therefore, we choose $g(\Hm) = \Rm$, where $g$ is the function mentioned in~\eqref{eq:g_function}. Note that other functions like the lower-triangular matrix from the $\mathbf{QL}$-decomposition are also possible. The fact that $\mathbf{QR}$-decomposition is also used in most sphere-decoding variants is the primary reason for our choice of $g$. Also, assuming that the used constellations are symmetric about the origin (as is the case with QAM which contains $-x$ for every constellation symbol $x$), we have 
\begin{equation}
\label{eq:diag_eqv1}
R_{\Dc,i}(\Rm) = R_{\Dc,i}(\Rm\Dm)
\end{equation}
where $\Dm$ is any $2N \times 2N$ sized diagonal matrix with the diagonal elements being $\pm 1$. From~\eqref{eq:eqv_signal_model}, it follows that for any orthogonal matrix $\Um \in \mathbb{R}^{2N\times 2N}$, we have $R_{\Dc,i}(\Rm) = R_{\Dc,i}(\Um\Rm)$. The only class of real-valued orthogonal matrices for which $\Um\Rm$ continues to be upper-triangular is that of diagonal matrices with $\pm 1$ entries. So, we have 
\begin{equation}
\label{eq:diag_eqv2}
R_{\Dc,i}(\Rm) = R_{\Dc,i}(\Dm\Rm)
\end{equation} 
where $\Dm$ is a $2N \times 2N$ sized diagonal matrix with the diagonal elements being $\pm 1$. These facts will be utilized during the training process for data augmentation, mainly to increase the robustness of training and also for increased speed of convergence.

Algorithm~\ref{alg:generating_labels} presents pseudocode to generate labeled data from a set of raw channels $\Hc$. In the algorithm, $N_{samp}$ is the number of samples used for Monte-Carlo approximation. The \gls{BMDR} approximations are computed over a set of useful transmit power values for each user. In particular, each channel matrix in the dataset is used to generate $N_p$ input feature-label pairs by drawing the transmit power values uniformly from $\LSB \rho_{min,i}, \rho_{max,i} \RSB$. The values of $\rho_{min,i}$ and $\rho_{min,i}$ are chosen such that the resulting \gls{BMDR} for each sample is between the minimum and maximum code-rate for that particular constellation $\Qc_i$. For example, in \gls{5GNR}, the minimum \gls{LDPC} code-rate is $0.11$ for $4-$QAM and the maximum is around $0.67$. So, the values of $\rho_{min,i}$ and $\rho_{min,i}$ should be chosen to obtain \gls{BMDR} labels within $\LB 0.1, 0.7 \RB$. The value of $N_p$ is dependent on $\rho_{max,i}- \rho_{min,i}$ since it is desirable to have a reasonable number of transmit power samples in the range $\LSB \rho_{min,i}, \rho_{max,i} \RSB$ per channel matrix. We found that $50$ is a good enough value through experimentation. If a practical channel estimation technique $CE$ is employed that outputs a post-processed channel estimate $\Hrm'$ of a true channel realization $\Hrm$, and a post-processed observed signal $\yrv'$ derived from the true observed signal $\yrv$, the update in Algorithm~\ref{alg:generating_labels} is given as 
\begin{align}
   r_{\Dc,i} \gets  r_{\Dc,i} +  \frac{\sum_{l=1}^{n_t^{(i)}}\sum_{j=1}^{m_i}\log{q_{\Dc,CE,i,l,j}(b_{i,l,j}; \yrv', \Hrm'_{norm}\Omega)}}{m_i n_t^{(i)}N_{samp}}, \forall i=1,\cdots, N_u.
\end{align}

\begin{figure}[t]
\centering
        \includegraphics[height = 0.25\textheight, width=0.8\textwidth]{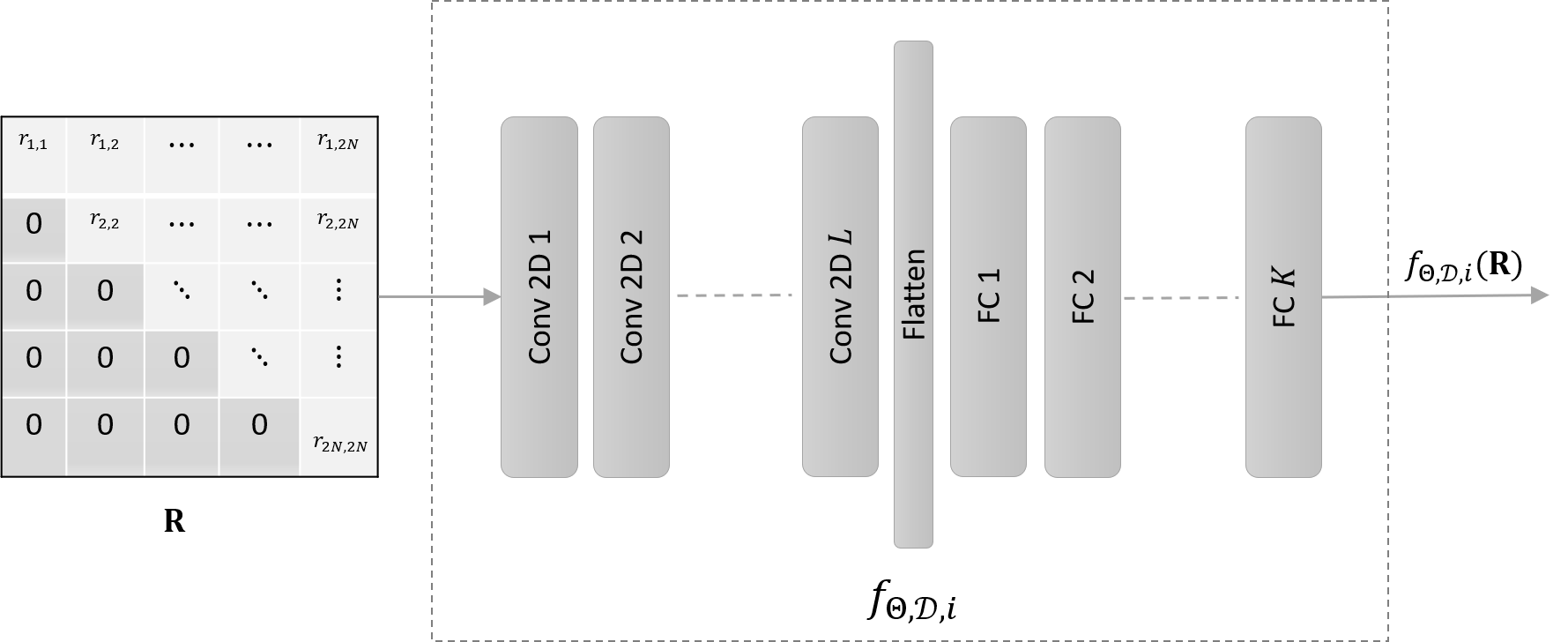}
    \caption{A general \gls{CNN} architecture for \gls{BMDR} prediction with $L$ \gls{Conv2D} layers and $K$ \gls{FC} layers.}
    \label{fig:cnn}
\end{figure}

\begin{algorithm}
\caption{Pseudocode for training a \gls{CNN} to perform \gls{BMDR} estimation for a detector $\Dc$ and \gls{UE} $i$, $i=1,\cdots,N_u$}
\label{alg:Training_cnn}
\begin{algorithmic}
  \Input
  \Desc{$\Xc_{train}$}{: Set of $n_{train}$ input features; is a subset of $\Xc_{L,\Dc}$ obtained in Algorithm \ref{alg:generating_labels}}
  \Desc{$\Yc_{train}$}{: Set of $n_{train}$ labels corresponding to $\Xc_{train}$; is a subset of $\Yc_{L, \Dc}^{(i)}$ }  
  \EndInput
  \Output
  \Desc{$f_{\Theta^*, \Dc, i}$}{: A trained \gls{CNN} with parameters $\Theta^*$ that predicts $R_{\Dc,i}(\Hm)$ given by~\eqref{eq:gen_bmdr_def} for an input $\Rm \in \Rbb^{2N \times 2N}$, the $\Rm$-matrix of the $\Qm\Rm$-decomposition of $\Hm^{\Rc}$}    
  \EndOutput 
\Initialize{\strut$\theta \gets \theta_{init}$, $\forall \theta \in \Theta$} 
\State $\Dbb \gets \LP \LB \Rrm^{(n)}, \bar{R}_{\Dc,i}^{(n)} \RB \middle | \Rrm^{(n)} \in \Xc_{train},   \bar{R}_{\Dc,i}^{(n)} \in \Yc_{train} \RP_{n=1}^{n_{train}}$ \Comment{Dataset of labeled pairs}
\While{Stopping Criteria {\it not reached}}
   \State $\{\Dbb_r\}_{r=1}^{B} \gets \mathrm{MiniBatch}\LB \Dbb \RB$ \Comment{Create $B$ random mini-batches}
   \For{$r \gets 1$ to $B$}
      \ForAll{$\LB \Rrm^{(n)}, \bar{R}_{\Dc,i}^{(n)} \RB \in \Dbb_r $}
        \State $\Lc_{\Theta,n}^{(1)} \gets \mathrm{Loss}\LB \bar{R}_{\Dc,i}^{(n)}, f_{\Theta, \Dc, i}\LB \Rrm^{(n)} \RB \RB$
        \State $d_k \gets \mathrm{SAMPLE}\LB \Uc (\{-1,1\}) \RB$, $\forall k = 1,\cdots, 2N$. 
        \State $\Drm \gets \mathrm{diag}\LB d_1,\cdots,d_{2N}\RB$  \Comment{Data-Augmentation}
        \State $\Lc_{\Theta,n}^{(2)} \gets \mathrm{Loss}\LB \bar{R}_{\Dc,i}^{(n)}, f_{\Theta, \Dc, i}\LB \Rrm^{(n)}\Drm  \RB \RB$ 
        \State $d_k \gets \mathrm{SAMPLE}\LB \Uc (\{-1,1\}) \RB$, $\forall k = 1,\cdots, 2N$. 
        \State $\Drm \gets \mathrm{diag}\LB d_1,\cdots,d_{2N} \RB$  \Comment{Data-Augmentation}
        \State $\Lc_{\Theta,n}^{(3)} \gets \mathrm{Loss}\LB \bar{R}_{\Dc,i}^{(n)}, f_{\Theta, \Dc, i}\LB \Drm\Rrm^{(n)} \RB \RB$
      \EndFor
      \State $\Lc_{\Theta} \gets \frac{1}{3\vert \Dbb_r \vert}\sum_{n=1}^{\vert \Dbb_r \vert}\sum_{l=1}^{3}\Lc_{\Theta,n}^{(l)}$
      \State $\nabla \Lc_{\Theta} \gets \mathrm{Gradients}(\Lc_{\Theta})$ \Comment{Compute gradients w.r.t each $\theta \in \Theta$}
      \State $\Theta \gets \mathrm{GradientDescentUpdate}\LB \Theta, \nabla \Lc_{\Theta} \RB$ \Comment{Update $\Theta$}
   \EndFor
\EndWhile  
\end{algorithmic}
\end{algorithm}

We use a \gls{CNN} for the purpose of \gls{BMDR} prediction. This is motivated by the belief that a \gls{CNN} has the ability to accurately identify the \gls{BMDR}-defining patterns in the $\Rm$-matrix. Fig.~\ref{fig:cnn} shows a general architecture of a simple \gls{CNN} used for our purpose. Assuming the signal model given by~\eqref{eq:signal_model_1}, the input to the \gls{CNN} is $\Rm$ where $\Rm \in \Rbb^{2N\times 2N}$ is obtained by the $\mathbf{QR}$-decomposition of $\Hm^{\Rc} \in \Rbb^{2n_r \times 2N}$. The pseudocode for training the \gls{CNN} is outlined in Algorithm~\ref{alg:Training_cnn}. As mentioned earlier, the observations in~\eqref{eq:diag_eqv1} and~\eqref{eq:diag_eqv2} are exploited for data-augmentation in order to increase the robustness of the trained model. The model in~\eqref{eq:g_function} indicates that the noise $\varepsilon$ is bounded, and within the bounded interval, we have observed from experiments that this noise is better approximated by a Laplacian random variable than a Gaussian one. Therefore, when the dataset does not have any zero-valued target labels, the loss function that we choose in Algorithm~\ref{alg:Training_cnn} for any single label $y_{true}$ and the corresponding predicted output ${y_{pred}}$ is given by 
\begin{equation}
\mathrm{Loss}(y_{true},y_{pred}) = \frac{\vert y_{true}- y_{pred} \vert}{\vert y_{true} \vert }
\end{equation}
so that the aggregated loss in a mini-batch is the {\it normalized mean absolute error}. Normalization aids in faster convergence. When the dataset has zero-valued target labels, we simply choose $
\mathrm{Loss}(y_{true},y_{pred}) = \vert y_{true}- y_{pred} \vert $ for the entire dataset so that the aggregated loss in a mini-batch is the mean absolute error.

\begin{note}
   In \gls{MU-MIMO} systems where the number of users served in each slot is variable, one can fix the maximum number of served users to be $N_u^{max}$, and use Algorithm~\ref{alg:generating_labels} to generate the labels by including $\rho_i = 0$ in the set of transmit powers levels for each $i=1,\cdots,N_u^{max}$.  
\end{note}
\begin{note}
   For non-linear detectors which are invariant to user permutation, i.e., the \gls{LLR} for each user is the same irrespective of the order of users, it is possible to use a single trained neural network for predicting the \gls{BMDR} for each user. However, the input to the neural network in order to predict a particular user's \gls{BMDR} needs to be the $\Rm$-matrix of the appropriately permuted channel matrix.  
\end{note}
\begin{note}
   \gls{5GNR} supports a maximum of $29$ \gls{MCS} levels and a maximum of four different QAM constellations \cite[Table 5.1.3.1-1--5.1.3.1-3]{3GPP_MCS_table_2020}. If the maximum number of users to be co-scheduled is $N_u^{max}$, the total number of \glspl{CNN} to be trained is $4^{N_u^{max}}$. In practice, assuming that each user is equipped with $2-4$ transmit antennas, $N_u^{max}$ is limited to $4$ or $5$ due to the limited availability of orthogonal \glspl{DMRS} which are $12$ in number \cite[Table 6.4.1.1.3-2]{3GPP_modulation_2020}. So, one might need to train at most $1024$ different \glspl{CNN} for practical implementation. We have observed that it is sufficient for each \gls{CNN} to be small-sized with around $5000$ parameters and that it takes around $5$ hours to complete both data preparation and training for each \gls{CNN}. Since the training is done offline, it takes around $210$ days to completely train all the $1024$ \glspl{CNN} sequentially on one machine. With parallel computation, the entire training procedure can be completed within a few days. By carefully choosing the training set of channel matrices to cover a wide range of condition numbers, the need to retrain the \glspl{CNN} is minimal.
\end{note}

\section{Simulation Results}
\label{sec:sim_results}

In this section, we present our evaluation setup and some simulation results in support of our claims. We consider the $32$-best detector as our choice of non-linear detector. Our reason for choosing $K=32$ is that it is easy to implement. We would like to point out that in practice, $K=64$ or higher is required for a significant improvement in performance compared to \gls{LMMSE}.

\subsection{Dataset and Training}

\begin{table}
	\begin{center}
	  \begin{tabular}{ | l | c | c | c | c | c | }
		\hline
		{\bf Layer}		& {\bf Channels/Neurons}	& {\bf Kernel size} & {\bf Activation} & {\bf Output Shape} &{ \bf Number of Parameters}	\\ \hline \hline
		Input Layer 	& N/A	& N/A		& N/A & $(B_S,8,8,1)$ & $0$ \\ \hline
		\gls{Conv2D} 1 	& $32$		& $(2,2)$	& $\ReLU$ & $(B_S,8,8,32)$  &	$160$ \\ \hline
		\gls{Conv2D} 2	& $16$	& $(2,2)$			& $\ReLU$		& $(B_S,8,8,16)$ &	$2064$	\\ \hline  
		\gls{Conv2D} 3	& $8$	& $(2,2)$			& $\ReLU$		& $(B_S,7,7,8)$ &	$520$	\\ \hline  
		Flatten	& N/A		& N/A		& 	N/A		& $(B_S,392)$ &	$0$ \\ \hline
		\gls{FC} 1	& $8$	& N/A	& $\ReLU$	& $(B_S,8)$ &	$3144$	\\ \hline   
		\gls{FC} 2	& $4$	& N/A	& $\ReLU$	& $(B_S,4)$ &	$36$	\\ \hline  
		\gls{FC} 3 	& $1$  	& N/A	& $\ReLU$	& $(B_S,1)$ &	$5$	\\ \hline
	  \end{tabular}
	\end{center}
	\caption{Architecture details of the CNN used for training, with a total of 5929 trainable parameters. Here $B_S$ denotes the batch size.}
	\label{tab:table1}
\end{table}

A dataset of channel realizations was generated for the 3GPP 38.901 UMa NLoS~\cite[Section 7.2]{3GPP_ch_model_2019} scenario at a carrier frequency of \SI{3.5}{\GHz} using the QuaDRiGa channel simulator. We considered a \gls{BS} equipped with a rectangular planar array consisting of $16$ ($2$ vertical, $8$ horizontal) single-polarized antennas installed at a height of \SI{25}{\metre} with an antenna spacing of $0.5 \lambda$, where $\lambda$ is the carrier wavelength. The \gls{BS} was assumed to have a coverage angle of \ang{120} and a coverage radius of \SI{250}{\metre}. Within this coverage sector, the region within a distance of \SI[input-quotient  = :, output-quotient = \text{--}]{25:250}{\metre} from the \gls{BS} was considered for the random placement of four single-antenna users ($N_u=4$). Once a user was dropped at a particular position, it was assumed to move along a linear trajectory with a speed varying between \SI[input-quotient  = :, output-quotient = \text{--}]{18:35}{\kmph}. Channels were sampled every $35.7$ \textmu s to obtain sequences of length $N_{symb}=70$ (corresponds to $5$ slots of $14$ symbols each in \gls{5GNR}). Each channel realization was then converted to the frequency domain assuming a bandwidth of \SI{2.16}{\MHz} and $N_{sc} = 72$ sub-carriers with a subcarrier spacing of \SI{30}{\kHz}, i.e., $6$ \glspl{PRB} in \gls{5GNR}. A total of $N_{drops} = 400$ independent user drops were obtained, resulting in $400*70*72 = 2.016$ million channel matrices of size $16\times 4$ arranged as an array of dimensions $(400,72,70,16,4)$, to be read as ($N_{drops}$, $N_{sc}$, $N_{symb}$, $n_r$, $N_u$). Since the path-loss can vary dramatically between different users, we assumed perfect power control per sequence of $70$ symbols so that
\begin{align}
\frac{1}{70N_{sc}}\sum_{f=1}^{N_{sc}}\sum_{t=1}^{70} \left\Vert \hrv_{f,t}^{(i)} \right\Vert^2 = 16, ~~\forall i=1,2,3,4,
\end{align}
where $\hrv_{f,t}^{(i)}$ is the channel from \gls{UE} $i$ to the \gls{BS} on subcarrier $f$ and symbol $t$. From the resulting set of matrices, we chose a set $\Hc$ of $10,000$ matrices with condition numbers uniformly distributed between \SI[input-quotient  = :, output-quotient = \text{--}]{0:25}{\dB}. For this set of channel matrices, a training dataset was generated for the $32$-best detector as described in Algorithm~\ref{alg:generating_labels} with $N_{samp} = 500$ and $N_p = 50$, resulting in $500,000$ feature-label pairs. The set of transmit powers (in \si{dB}) was generated by randomly sampling from $[-16, -6]$, $[-8, 0]$, and $[-4, 10]$, respectively, for $4/16/64-$QAM. These ranges were chosen so that for at least one detector, the \gls{BMDR} value (label) for any data sample was at least $0.1$ for all users. In total, there are $3^4 = 81$ combinations of user constellations and $500,000$ feature-label pairs for each of them. A test dataset was generated similarly for a separate set of $5000$ channel matrices. The standard deviations of the test labels are $0.25$ and $0.27$ for the \gls{LMMSE} detector and the $32$-best detector, respectively. So, this dataset contains input features with a wide variety of label values.	 Table~\ref{tab:table1} details the architecture of the \gls{CNN} that was trained using Algorithm~\ref{alg:Training_cnn} to predict \gls{BMDR} for the $32$-best detector. 

\subsection{Numerical Results}

We used the trained \gls{CNN} to predict the BMDR on the test data with $250,000$ feature-label pairs for each of the $81$ combinations of QAM constellations for the four users. As for the LMMSE detector, the labels were generated as explained in Algorithm~\ref{alg:generating_labels}, but the prediction was performed as explained in Remark~\ref{rem:lmmse_pred}.

\begin{figure}
	\centering			
		\includegraphics[scale = 0.5]{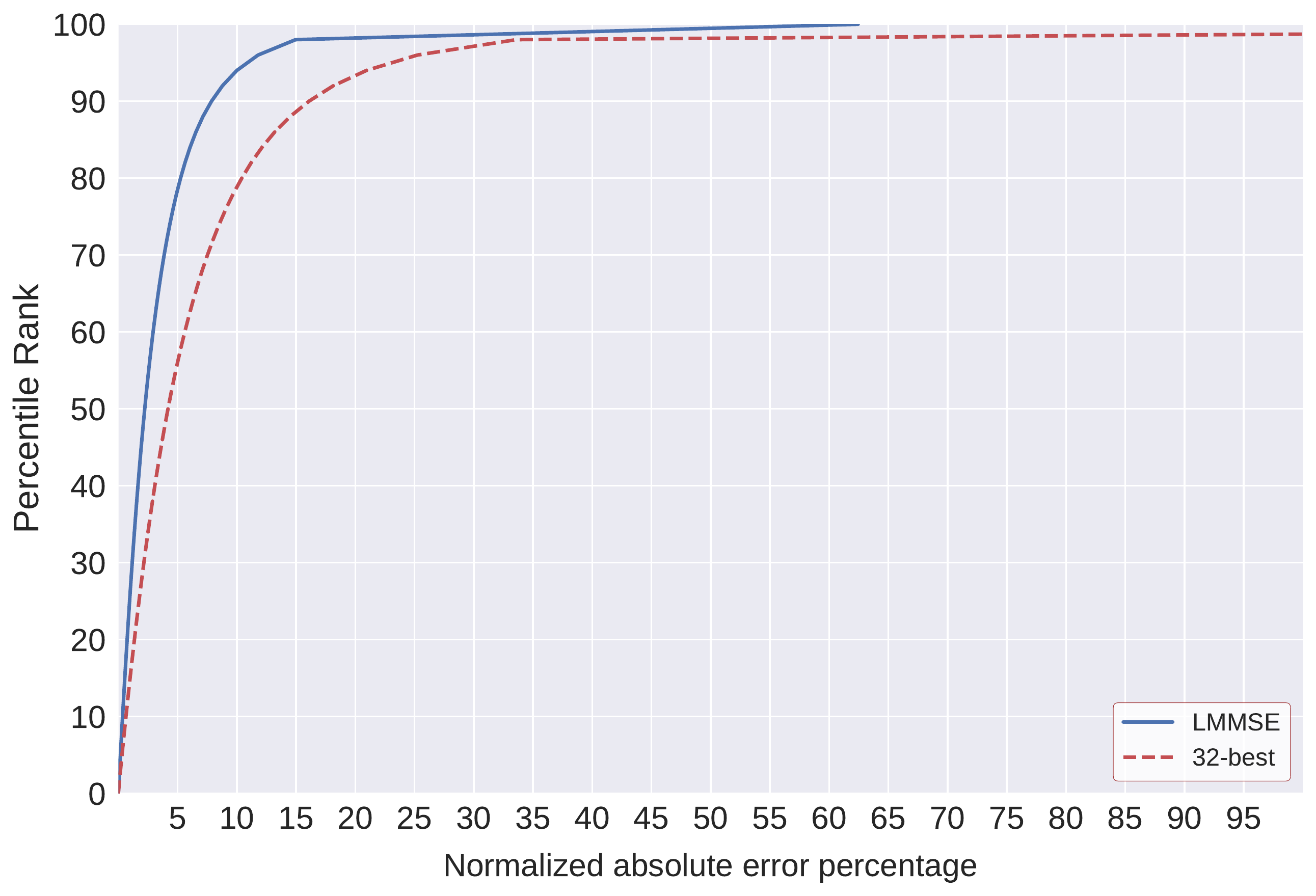}
		\caption{The percentiles of the normalized absolute BMDR prediction error for the \gls{LMMSE} and $32$-best detectors.}
		\label{fig:lmmse_error_percentile}
\end{figure}

Fig.~\ref{fig:lmmse_error_percentile} shows the plot of the percentile ranks of the normalized absolute value of the prediction error, i.e., $\vert \varepsilon \vert /\bar{R}_{\Dc,i}(\Hm) $ for $\varepsilon$ and $\bar{R}_{\Dc,i}(\Hm)$ as given in~\eqref{eq:g_function}. The percentiles here refer to the percentage of test predictions that have normalized absolute error percentages less than any given value on the X-axis. The accuracy of prediction for the $32$-best detector is slightly worse compared to that of the LMMSE detector. This is expected, since a non-linear detector will exhibit more variations in \gls{BMDR} compared to a linear detector for the same amount of perturbation in the channel coefficients. However, most of the inaccuracies for the $32$-best detector occur at extremely low values of the labels. In particular, there are many data samples for which the \gls{BMDR} of the \gls{LMMSE} detector is at least $0.1$ but that for the $32$-best detector is less than $0.1$, and the prediction for these samples account for the large deviations in the case of the $32$-best detector. Considering that the standard deviation of the test labels is $0.27$ for the $32$-best detector, the prediction results are quite promising. 

\begin{figure}
	\centering			
		\includegraphics[scale = 0.48]{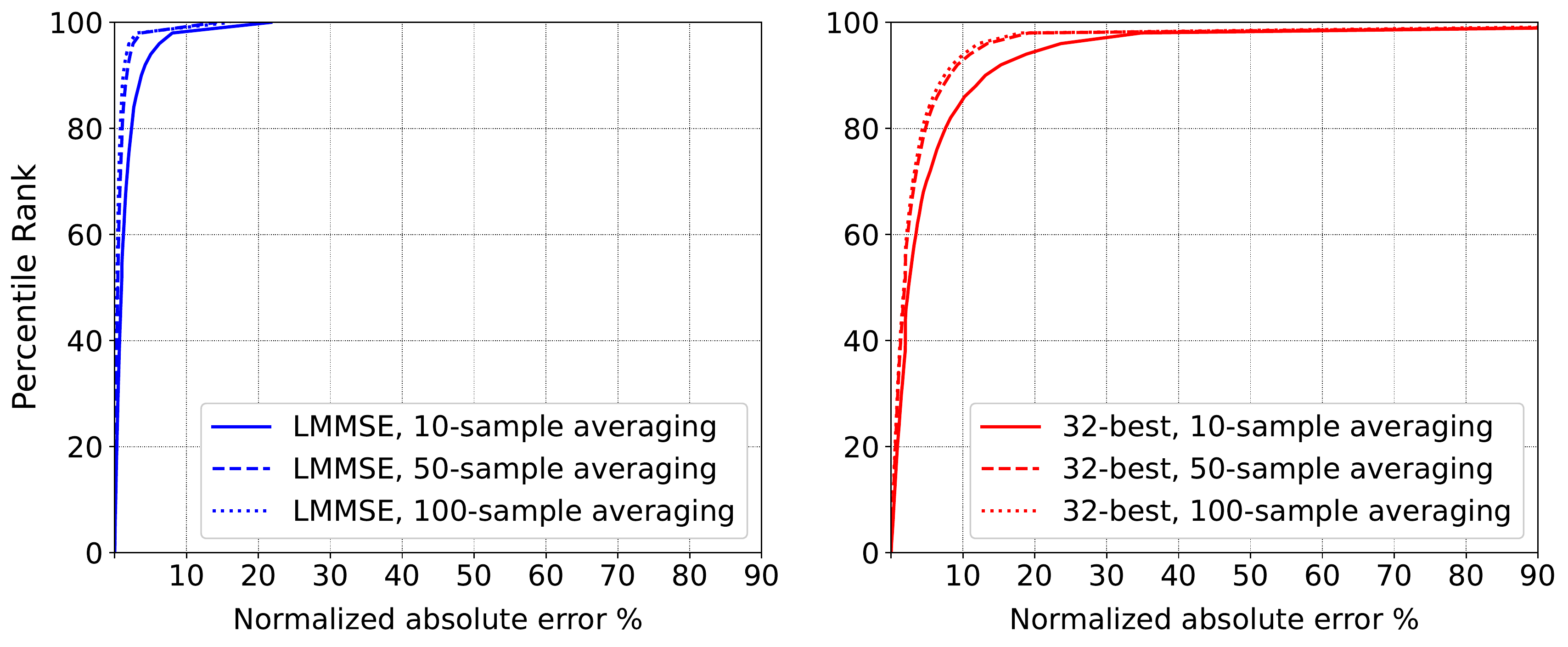}		
		\caption{The percentiles of the normalized absolute error with averaging over different sequence lengths for the LMMSE detector (left) and the $32$-best detector (right).}
		\label{fig:kbest_lmmse_avg_error_percentile}
\end{figure}

In real-world applications, we do not actually rely on \gls{BMDR} prediction for one channel realization, but rather on a sequence of channel realizations over which the codeword bits are transmitted. Therefore, we are more interested in the \gls{BMDR}-prediction errors for sequences of channel realizations. In view of this, we considered sequences of channel realizations of length $n_{seq} \in \LP10, 50, 100 \RP$ in the test dataset, each sequence consisting of channel realizations for consecutive \glspl{RE} in frequency. For each value of $n_{seq}$, we considered $20,000$ independent sequences with the start of each sequence being arbitrary, and recorded the normalized prediction errors of the \gls{BMDR} average for each of them. In particular, if the prediction error for sample $k$ with channel matrix $\Hrm_k$ in a sequence was $\varepsilon_k$ and the true value of the label was $\bar{R}_{\Dc,i}(\Hrm_k)$, the normalized prediction error of the average was calculated to be $\vert \sum_{k=1}^{n_{seq}} \varepsilon_k\vert/\sum_{k=1}^{n_{seq}} \bar{R}_{\Dc,i}(\Hrm_k)$. The percentiles of this normalized error are plotted in Fig. \ref{fig:kbest_lmmse_avg_error_percentile} for both the detectors. It can be seen that the normalized prediction error with averaging is more accurate than that for a single channel realization even for the $32$-best detector, and this is promising from the perspective of both \gls{LA} and \gls{PHY} abstraction. 

\section{Discussion and Concluding Remarks}
\label{sec:conc_remarks}

For \gls{MU-MIMO} systems with non-linear receivers, \gls{LA} and \gls{PHY} abstraction are quite challenging to perform due to the absence of a suitable metric that is equivalent to post-equalization \gls{SINR} in linear receivers. In this paper, we introduced \gls{BMDR} of a detector as one such equivalent metric, and connected it to existing information-theoretic results on \gls{GMI} for \gls{BICM} systems with possibly mismatched decoding. The relationship between \gls{BMDR}, transmission code-rate, and \gls{CER} for a detector was established, and a method to estimate the BMDR using a trained \gls{CNN} was presented. Simulation results were provided to substantiate our claims. 

While the training process itself is quick and the objective of BMDR prediction can be achieved with relatively simple \gls{CNN} models, the task that involves the highest amount of computational resources is the preparation of labeled data. For every \gls{MU-MIMO} configuration, defined by the number of receive antennas $n_r$, number of users $N_u$, number of transmit antennas $n_t^{(i)}$ for \gls{UE} $i$, and constellation $\Qc_i$ for \gls{UE} $i$, one needs to prepare labeled data. So, this would probably limit the usage of the proposed technique to relatively smaller \gls{MU-MIMO} configurations with about $4-5$ co-scheduled users. But the training is done offline, and the requirement to retrain a properly trained \gls{CNN} is minimal. This makes the proposed \gls{BMDR} prediction technique practical for commercial applications. In the next part of this paper, the concepts developed thus far will be further exploited to perform \gls{LA} and \gls{PHY} abstraction for both linear and non-linear \gls{MU-MIMO} receivers.

\begin{appendices}

\section*{Appendix}

\subsection*{Proof of Theorem~\ref{thm:main_thm}}

To prove the achievability part, we first suppose that the set of channel realizations $\Hc_{\Gc_i(n)}$ is held fixed. Noting that $n_i = nm_in_t^{(i)}$ and that $q_{\Dc}\LB \boldsymbol{b}_{f,t,i,l,j}; \yv_{f,t}, \Hrm_{f,t}\RB$ as defined by \eqref{eq:qd} represents the possibly mismatched posterior distribution of $\boldsymbol{b}_{f,t,i,l,j}$ when detector $\Dc$ is used, we plug $s=1$ into \eqref{eq:gmi_s} to obtain
\begin{align}
I_{gmi} & \geq \frac{1}{n_i}\sum_{(f,t)\in \Gc_i(n)}\Expect{\yv_{f,t}}{\sum_{l=1}^{n_t^{(i)}} \sum_{j=1}^{m_i}\log{\frac{q_{\Dc}\LB \boldsymbol{b}_{f,t,i,l,j}; \yv_{f,t}, \Hrm_{f,t} \RB}{\frac{1}{2}}}} \\
& = 1 + \frac{1}{n}\sum_{(f,t)\in \Gc_i(n)}\Expect{\yv_{f,t}}{\frac{1}{m_in_t^{(i)}}\sum_{l=1}^{n_t^{(i)}} \sum_{j=1}^{m_i}\log{q_{\Dc}\LB \boldsymbol{b}_{f,t,i,l,j}; \yv_{f,t}, \Hrm_{f,t} \RB}} \geq  R_{\Dc,i}^{min}.
\end{align}
Since $I_{gmi}$ is an achievable rate, so is $R_{\Dc,i}^{min}$.

\begin{figure}[t]
    \centering
            \includegraphics[width=\textwidth]{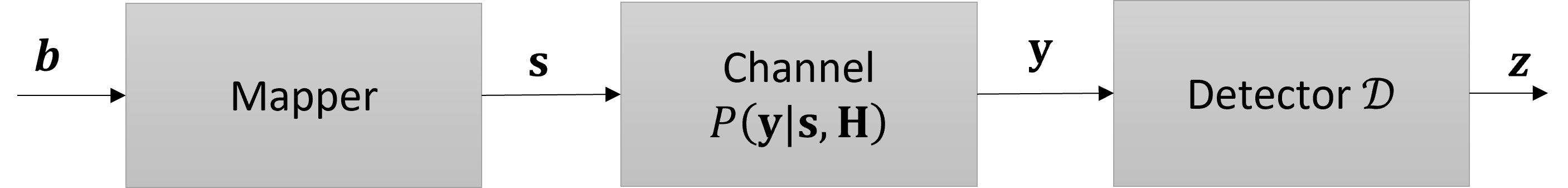}
        \caption{\gls{BICM} system with possibly mismatched decoding.}
        \label{fig:mutual_information}
\end{figure}

To prove the other part of the theorem, we proceed as follows. Let $\boldsymbol{z}_{f,t}$ denote the symbol at the output of the detector with $\boldsymbol{z}_{f,t} = f\LB \yv_{f,t}\vert \Hm_{f,t}\RB$ for some deterministic function $f$, and let $\Zbc_{\Gc_i(n)} \triangleq \{ \boldsymbol{z}_{f,t}, \forall (f,t) \in \Gc_i(n)\}$. Further, let $I(X;Y)$ denotes the \gls{MI} between random variables $X$ and $Y$. Since the $\{\yv_{f,t}\}$ are independent and $\boldsymbol{z}_{f,t}$ is a deterministic function of $\yv_{f,t}$, we have
\begin{align} \label{eq:main_MI}
    I \LB \Bbc_{\Gc_i(n)}; \Zbc_{\Gc_i(n)}|\Hbc_{\Gc_i(n)}= \Hc_{\Gc_i(n)} \RB & = \sum_{(f,t)\in \Gc_i(n)}I \LB\{\boldsymbol{b}_{f,t,i,l,j}\}_{j,l}; \boldsymbol{z}_{f,t}|\Hm_{f,t} = \Hrm_{f,t} \RB \\ \label{eq:main_MI_1}
    &= \sum_{(f,t)\in \Gc_i(n)}\sum_{j,l}I(\boldsymbol{b}_{f,t,i,l,j}; \boldsymbol{z}_{f,t}|\Hm_{f,t} = \Hrm_{f,t})
\end{align}
 where \eqref{eq:main_MI_1} is due to the assumption that $\LP \boldsymbol{b}_{f,t,i,l,j}, \forall l =1,\cdots,n_t^{(i)}, \forall j=1,\cdots, m_i\RP$ are independent due to interleaving (the classical \gls{BICM} system).  For convenience, we drop the subscripts and use the notation as shown in Fig.~\ref{fig:mutual_information}. It can be seen that the mutual information between $\boldsymbol{b}$ and $\boldsymbol{z}$ for a channel realization $\Hrm$ is 
\begin{align}
I(\boldsymbol{b}; \boldsymbol{z} \vert \Hm = \Hrm) & = \Expect{\boldsymbol{b},\boldsymbol{z} \vert \Hm}{\log{\frac{\Prob{\boldsymbol{b} \vert \boldsymbol{z}, \Hm = \Hrm}}{\Prob{\boldsymbol{b}}}}} \\
& = 1 + \Expect{\boldsymbol{b},\boldsymbol{z}\vert \Hm}{\log{\Prob{\boldsymbol{b} \vert \boldsymbol{z},\Hm = \Hrm}}}
\end{align}
where $\Prob{\boldsymbol{b}} = 1/2$ for $ \boldsymbol{b} \in \{0,1\}$, and $\Prob{\boldsymbol{b} \vert \boldsymbol{z},\Hm = \Hrm}$ denotes the posterior distribution of $\boldsymbol{b}$ conditioned on $\boldsymbol{z}$ and $\Hm = \Hrm$. By definition, $q^*_{\Dc}(b_{f,t,i,l,j}; \yrv_{f,t}, \Hrm_{f,t})$ as given by \eqref{q_def} is the posterior probability of a transmitted bit $b_{f,t,i,l,j}$ given that a detector $\Dc$ is used. Therefore, 
\begin{align}
    I(\boldsymbol{b}_{f,t,i,l,j}; \boldsymbol{z}_{f,t}|\Hm_{f,t} = \Hrm_{f,t})  & = 1 + \Expect{\boldsymbol{b}_{f,t,i,l,j} , \boldsymbol{z}_{f,t} \vert \Hm_{f,t}}{\log{q^*_{\Dc}\LB \boldsymbol{b}_{f,t,i,l,j}; \yv_{f,t}, \Hrm_{f,t} \RB}} \\ \label{eq:step1}
    & = 1 + \Expect{\yv_{f,t} \vert \Hm_{f,t}}{\log{q^*_{\Dc}\LB \boldsymbol{b}_{f,t,i,l,j}; \yv_{f,t}, \Hrm_{f,t} \RB}},
\end{align}
where \eqref{eq:step1} is due to the fact that $\boldsymbol{z}_{f,t}$ is a deterministic function of $\yv_{f,t}$. Using~\eqref{eq:step1} in~\eqref{eq:main_MI}, we obtain
\begin{align}
I(\Bbc_{\Gc_i(n)};\Zbc_{\Gc_i(n)}|\Hbc_{\Gc_i(n)}= \Hc_{\Gc_i(n)}) & =  \sum_{(f,t)\in \Gc_i}\sum_{j,l}I(\boldsymbol{b}_{f,t,i,l,j}; \boldsymbol{z}_{f,t}|\Hm_{f,t} = \Hrm_{f,t}) \\
& = n_iR^*_{\Dc,i}(\Hc_{\Gc_i(n)}) \leq n_iR_{\Dc,i}^{max^*}.
\end{align}

Here, $R^*_{\Dc,i}(\Hc_{\Gc_i(n)})$ represents the normalized (by the codeword length) mutual information between the transmitted bits and the detector output conditioned on a set of channel realizations $\Hc_{\Gc_i(n)}$. It can be viewed as the \gls{BICM} capacity (with the channels kept fixed) when an arbitrary detector $\Dc$ is used (as opposed to an optimal detector).  So, there exists no code which can transmit at a rate higher than it for reliable transmission. In fact, $R^*_{\Dc,i}(\Hc_{\Gc_i(n)})$ is also achievable with ideal \gls{LLR} correction provided that $\Hc_{\Gc_i(n)}$ is held fixed. This is seen by plugging in $s=1$ in \eqref{eq:gmi_s} for $q^*_{\Dc}$. This has also been shown in \cite[Theorem 1]{Jalden2010}.
\end{appendices}

\bibliographystyle{IEEEtran}
\bibliography{IEEEabrv, references}

\end{document}